\documentclass[sn-mathphys-num,iicol]{sn-jnl}

\usepackage{graphicx}%
\usepackage{multirow}%
\usepackage{amsmath,amssymb,amsfonts}%
\usepackage{amsthm}%
\usepackage{mathrsfs}%
\usepackage{xcolor}%
\usepackage{textcomp}%
\usepackage{manyfoot}%
\usepackage{booktabs}%
\usepackage{algorithm}%
\usepackage{algorithmicx}%
\usepackage{algpseudocode}%
\usepackage{listings}%

\usepackage[separate-uncertainty = true,multi-part-units=single]{siunitx}
\DeclareSIUnit\clight{\text{\ensuremath{c}}}
\usepackage{subcaption}
\usepackage{comment}
\usepackage{todonotes}
\usepackage[switch]{lineno} 
\usepackage{lmodern}

\begin{document}
\title[Article Title]{A Data-Driven Fast Simulation Approach for MAPS-based Detectors and their Optimization}

\author*[1]{\fnm{Dumitru Vlad} \sur{Berlea}}\email{vlad.berlea@desy.de}
\author*[1,2]{\fnm{Lucian} \sur{Fasselt}}\email{lucian.fasselt@desy.de}
\author[3]{\fnm{Prafulla} \sur{Behera}}
\author[4]{\fnm{Daniela} \sur{Bortoletto}}
\author[5]{\fnm{Craig} \sur{Buttar}}
\author[3]{\fnm{Theertha} \sur{Chembakan}}
\author[6]{\fnm{Valerio} \sur{Dao}}
\author[3]{\fnm{Ganapati} \sur{Dash}}
\author[7]{\fnm{Sebastian} \sur{Haberl}}
\author[7]{\fnm{Tomohiro} \sur{Inada}}
\author[8]{\fnm{Fuat Kerem} \sur{Isik}}
\author[1,2]{\fnm{Cigdem} \sur{Issever}}
\author[9]{\fnm{Xuan} \sur{Li}}
\author[10]{\fnm{Long} \sur{Li}}
\author[7]{\fnm{Heinz} \sur{Pernegger}}
\author[7]{\fnm{Petra} \sur{Riedler}}
\author[7]{\fnm{Walter} \sur{Snoeys}}
\author[7]{\fnm{Carlos} \sur{Solans Sánchez}}
\author[7]{\fnm{Anna} \sur{Swoboda}}
\author[8]{\fnm{Ilkay} \sur{Turk Cakir}}
\author[7]{\fnm{Milou} \sur{van Rijnbach}}
\author[3]{\fnm{Anusree} \sur{Vijay}}
\author[7]{\fnm{Julian} \sur{Weick}}
\author[1,2]{\fnm{Steven} \sur{Worm}}

\affil[1]{\orgname{Deutsches Elektronen-Synchrotron DESY}, \orgaddress{\city{Zeuthen}, \country{Germany}}}
\affil[2]{\orgname{Humboldt University of Berlin}, \orgaddress{\city{Berlin}, \country{Germany}}}
\affil[3]{\orgname{Indian Institute of Technology Madras}, \orgaddress{\city{Chennai}, \country{India}}}
\affil[4]{\orgname{University of Oxford}, \orgaddress{\city{Oxford}, \country{UK}}}
\affil[5]{\orgname{University of Glasgow}, \orgaddress{\city{Glasgow}, \country{UK}}}
\affil[6]{\orgname{Stony Brook University}, \orgaddress{\city{New York}, \country{USA}}}
\affil[7]{\orgname{CERN}, \orgaddress{\city{Geneva}, \country{Switzerland}}}
\affil[8]{\orgname{University of Ankara}, \orgaddress{\city{Ankara}, \country{Turkey}}}
\affil[9]{\orgname{Los Alamos National Laboratory}, \orgaddress{\city{Los Alamos}, \country{USA}}}
\affil[10]{\orgname{University of Birmingham}, \orgaddress{\city{Birmingham}, \country{UK}}} 

\abstract{A parametric simulation tool for pixel sensors is presented.
A realistic pixel response is simulated purely based on measurement input, without requiring detailed knowledge of the underlying manufacturing process.
As such, it provides an efficient alternative to the use of Technology Computer-Aided Design simulations, which typically depend on proprietary process information.
Due to its parametric approach, the package is fast and thus particularly useful for larger detector systems and high hit rate environments.
This work presents measurements, simulation and its validation for the MALTA2 sensor.
It is a small collection electrode monolithic active pixel sensor produced in the Tower \SI{180}{\nano\meter} Complementary Metal-Oxide-Semiconductor imaging process.
Modifications to the sensor's periphery, mainly in the hit merger, are studied in order to optimize the performance for tracking and calorimetry.
This optimization is of special interest as part of the MALTA3 sensor redesign in the \SI{65}{\nano\meter} Tower Partners Semiconductor Co. process.}


\maketitle
\section{Introduction}
The accurate simulation of silicon sensors is essential for the design of next-generation particle physics detectors.
This is particularly important for detector components close to the beam pipe of accelerator experiments, such as tracking or vertexing detectors, due to their strict detection requirements. Currently, the most powerful and accurate simulation tools used for optimizing semiconductor detector design are based on Technology Computer-Aided Design (TCAD). Such simulations are the cornerstone for optimizing analog chip design for charge imaging applications. 
Despite their predictive power, TCAD approaches rely heavily on detailed knowledge of silicon processing parameters, which are often proprietary to the manufacturer or require parameter tuning to data~\cite{Munker_2019,WENNLOF2025170227}. 

In this paper, we present a parametric simulation technique for complex silicon devices. It does not rely on chip design information, but rather on a parametrization of charge sharing and timing information using test beam and laboratory measurements described in Sections \ref{Sec:2.2} and \ref{Sec:2.3}. This offers an efficient simulation of the sensor performance with predictive power for the digital design of the read-out. Such an approach is well suited for simulations of Monolithic Active Pixel Sensors (MAPS), where pixel level digitization is often implemented. The simulation is structured as a GEANT4 suite and C++ analysis package which can be imported and reused for various applications. 
Currently, the package is considered for several particle physics experiments such as the ATLAS Inner Tracker upgrade at the High-Luminosity LHC \cite{ITK}, the Fast MAPS timing detector for the ePIC experiment at the EIC \cite{EIC} and position sensitive layers in digital calorimeters.

The fast data driven simulation technique is considered for two applications.
First, it enables the optimization of large-scale detector geometries for both tracking and calorimetry applications.
Second, the digital read-out can be optimized for future redesigns of MAPS. 
The latter is investigated in this work, with preliminary suggestions for optimal sensor design being made for the upcoming redesign of the MALTA3 sensor in the \SI{65}{\nano\meter} TPSCo technology node. The sensor will aim to satisfy both tracking and digital calorimetry applications.

Such a simulation could additionally be generalized for other sensors as long as data sets equivalent to what is presented in the following section are acquired.

\section{Charge collection measurement}
\subsection{The MALTA2 sensor}

The MALTA2 sensor is the second full-scale prototype of the MALTA sensor family. It is a fully Depleted Monolithic Active Pixel Sensor (DMAPS) produced in the Tower \SI{180}{\nano\meter} CMOS Imaging Process. 
It features a $224\times512$ pixel array with a total active area of $9\times$\SI{18}{\milli\meter^2}. Each pixel has a symmetric $36.4\times$\SI{36.4}{\micro\meter^2} pitch, with a small octagonal collection electrode (\SI{2}{\micro\meter} diameter). MALTA features a fully asynchronous read-out which does not require a clock propagation across the pixel array, leading to optimized power dissipation \cite{FPiroFE} and timing performance \cite{Gustavino_2022}. The sensor read-out architecture consists of $2\times8$ pixel groups that distribute data packets ($40$ bit words) across double columns.

\begin{figure}
\centering
\includegraphics[width=\linewidth]{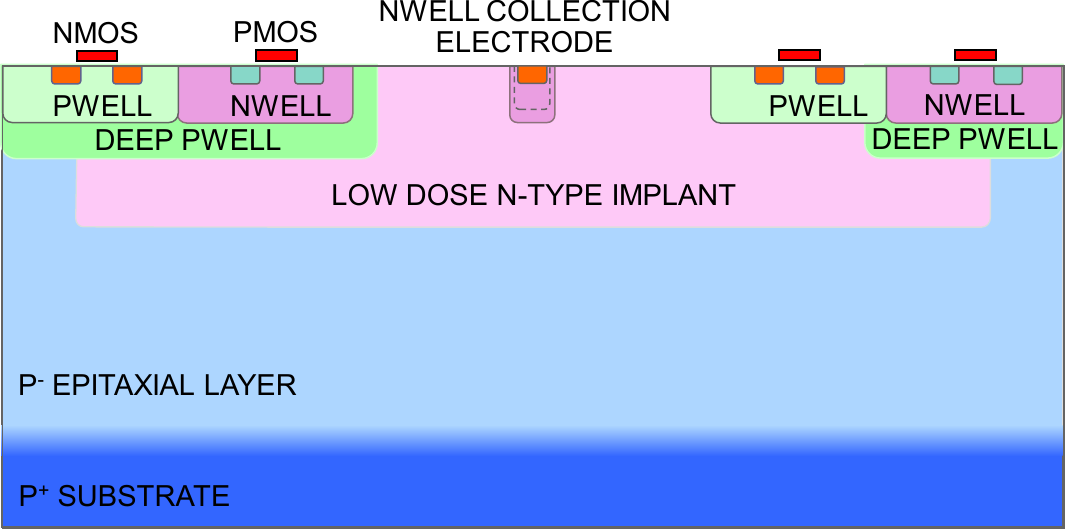}
\caption{MALTA2 pixel cross-section on \SI{30}{\micro\meter} thick epitaxial silicon with a gap in the low dose n-type implant. The features of this illustration are not to scale.}\label{fig:PixelCrosssection}
\end{figure}

Several analog process modifications have been investigated over the years in order to primarily optimize the sensor's radiation hardness \cite{Dyndal_2020,BDVRadFE}. In this paper, the MALTA2 sample has an epitaxial active volume grown on top of low resistivity silicon, which ensures the mechanical stability of the sensor. The process implements a low dose implant underneath the small collection electrode, which laterally expands the depletion volume across the pixel. An additional process modification adds a gap in the low dose implant, in order to optimize the pixel corner charge collection. 
Figure \ref{fig:PixelCrosssection} shows a schematic of the pixel cross-section.

\subsection{Charge sharing model}
\label{Sec:2.2}
\begin{figure*} 
\centering
\begin{subfigure}[t]{0.49\textwidth} 
\centering
\includegraphics[width=\linewidth]{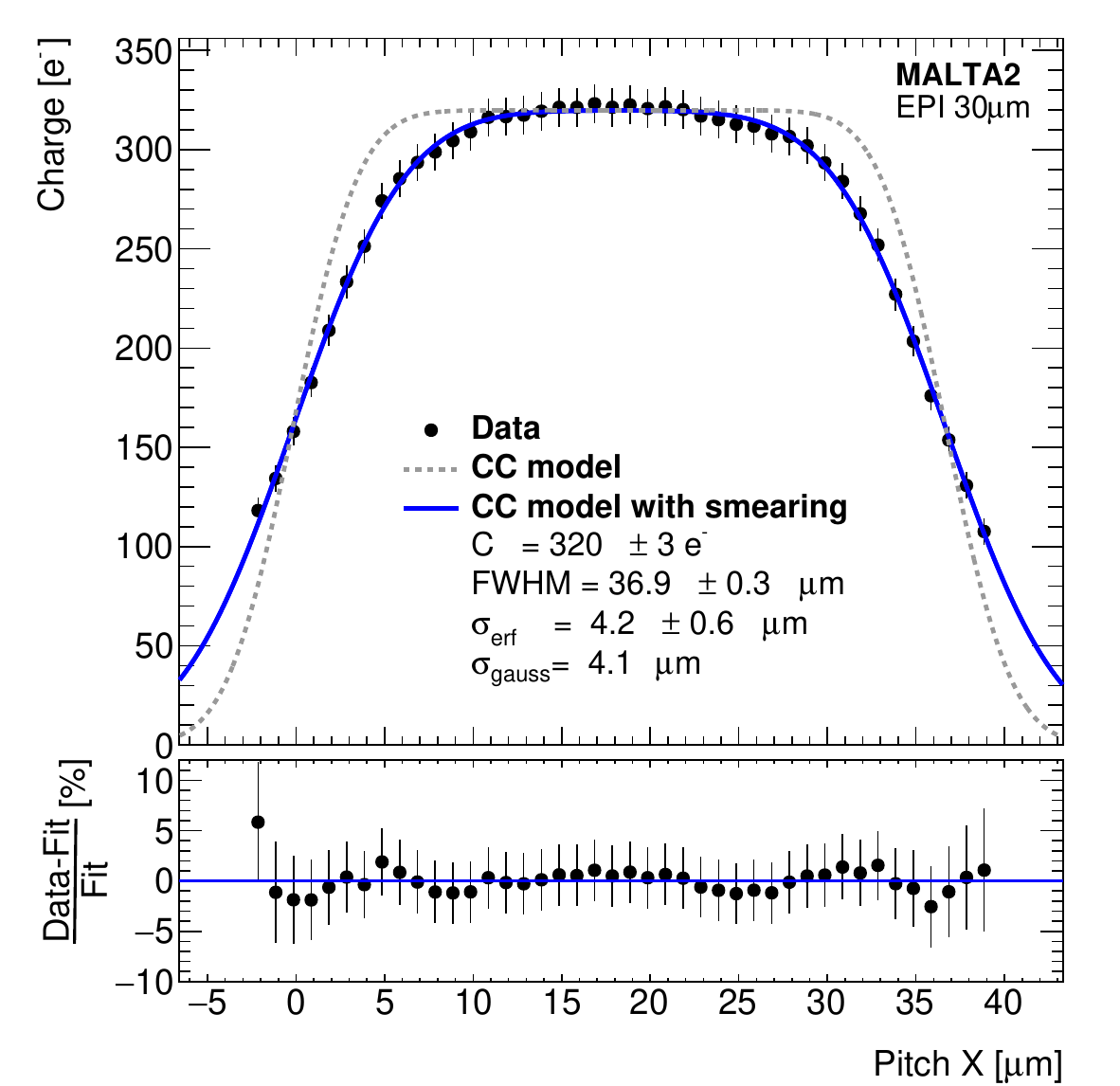}
\caption{One-dimensional pixel scan via E-TCT.}\label{fig:1DModel_TCT}
\end{subfigure}
\begin{subfigure}[t]{0.49\textwidth}
\centering
\includegraphics[width=\linewidth]{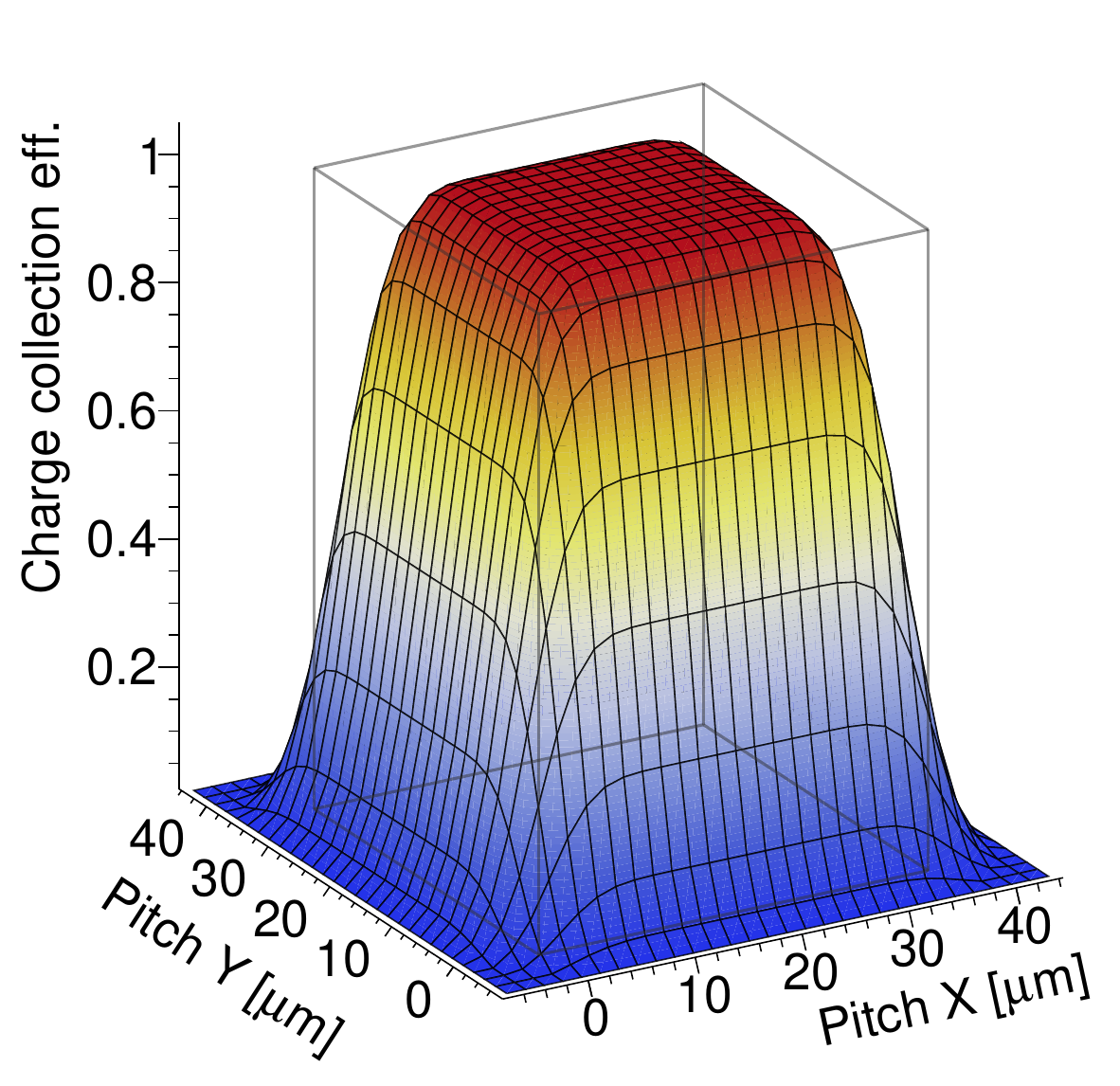}
\caption{Two-dimensional charge collection efficiency reconstructed from test beam data.}\label{fig:2DCCModel}
\end{subfigure}
\caption{Charge collection model derived from E-TCT and test beam data.
In (a) the charge is measured across the pixel pitch by scanning with an infrared laser.
The charge collection model (dashed line) describes the data when convoluted with the beam width $\sigma_{\mathrm{gauss}}$ (blue line).
The error-bars are dominated by the charge calibration of \SI{3}{\percent}.
A two-dimensional charge collection model as illustrated in (b) is obtained by charge reconstruction from test beam and is compatible with the E-TCT results~\cite{FASSELT_Parameterization}.
The gray box illustrates the nominal size of the $36.4\times\SI{36.4}{\square\micro\meter}$ pixel.
}\label{fig:CCModel}
\end{figure*}

The parametric simulation presented in this paper is based on a charge sharing model derived from data, obtained via the edge transient-current technique (E-TCT) and test beam at the CERN SPS.
Figure~\ref{fig:1DModel_TCT} shows E-TCT data acquired with a \SI{1064}{\nano\meter} laser by scanning the polished edge of a MALTA2 pixel.
The charge amplitude is reconstructed from digital data~\cite{FASSELTCHRECO,FASSELTCHCAL}.
The charge collection model (gray dashed line) is defined as a function of the position $x$ within the pixel by:
\begin{equation}
\label{eq:CC}
\mathrm{CC}(x;C,\sigma_{\mathrm{erf}}) = \frac{C}{2}\left[\mathrm{erf}\left(\frac{x}{\sigma_{\mathrm{erf}}}\right)-\mathrm{erf}\left(\frac{x-\mathrm{pitch}}{\sigma_{\mathrm{erf}}}\right)\right] 
\end{equation}
$C$ is the normalization constant and $\sigma_{\mathrm{erf}}$ is the width of the error-function that describes the charge collection at the pixel edges.
The model is convoluted with a Gaussian distribution, which accounts for the smearing due to the width of the laser beam $\sigma_{\mathrm{gauss}} = \SI{4.1}{\micro\meter}$~\cite{Berlea:2024TCT}.
The convolution (blue line) is fitted to the data.
A value of $\sigma_{\mathrm{erf}} = 4.2\pm\SI{0.6}{\micro\meter}$ is obtained for the width parameter of the charge collection model.
The measurement on the sensor edge gives access to only one lateral axis (pitch), while the other lateral axis is parallel to the incident laser beam.
To test the symmetry of the charge collection, a MALTA2 sensor on epitaxial silicon was also characterized at the CERN SPS with a \SI[per-mode=symbol]{120}{\giga\electronvolt\per\clight} charged hadron beam.
In this setup, the beam was incident on the top of the sensor.
The charge deposition follows a Landau-distribution.
Its most probable value is reconstructed and serves as input for a two-dimensional charge collection model~\cite{FASSELT_Parameterization}.
This is obtained by multiplying two one-dimensional functions:
\begin{equation}
\label{eq:CC_2D}
\mathrm{CC_{2D}}(x,y;C,\sigma_{\mathrm{erf}}) = \mathrm{CC}(x;C,\sigma_{\mathrm{erf}}) \times \mathrm{CC}(y;1,\sigma_{\mathrm{erf}}). 
\end{equation}
As shown in Figure~\ref{fig:2DCCModel}, a model symmetric in $x$ and $y$ is obtained and described by:
\begin{equation}
    \sigma_{\mathrm{erf}} = 4.3\pm\SI{0.3}{\micro\meter}
\end{equation}
This parameter is compatible with the complementary E-TCT result and serves as the main input for the charge sharing model.
In the simulation, deposited charge generated by \textsc{Geant4}~\cite{Geant4} is distributed among the pixel closest to the ionization step (the seed pixel) and the three pixels adjacent to the nearest pixel corner.
For a charge deposited at position $(x,y)$, the fraction assigned to the seed pixel is given by
$\mathrm{CC_{2D}}(x,y; C=1, \sigma_{\mathrm{erf}}=\SI{4.3}{\micro\meter})$.
The charge fractions for the neighboring pixels are obtained by evaluating $\mathrm{CC_{2D}}$ at positions where $x$ and/or $y$ are shifted by one pixel pitch.
By construction, the four resulting fractions sum to unity, neglecting any charge loss effects because the sensitive region is expected to be fully depleted.
This approach restricts charge sharing from a single energy deposition to pixel clusters of at most four pixels.
Larger clusters may nevertheless arise from multiple ionization processes across different pixels, for example those induced by delta rays.
Overall, the model is computationally lightweight, as signal formation only requires the evaluation of a parametric expression.
The description of charge collection can be further generalized by including: sharing over more than four pixels; asymmetric behavior by introducing different values for $\sigma_{\mathrm{erf,x}}$ and $\sigma_{\mathrm{erf,y}}$; or implementing an explicit dependence on the $z$ coordinate.
For MALTA2 the accurate description of the cluster size is validated in Section~\ref{sec:ClSizeValid}.
 
\subsection{Time walk}
\label{Sec:2.3}
\begin{figure*} 
\centering
\begin{subfigure}[t]{0.49\textwidth}
\centering
\includegraphics[width=\linewidth]{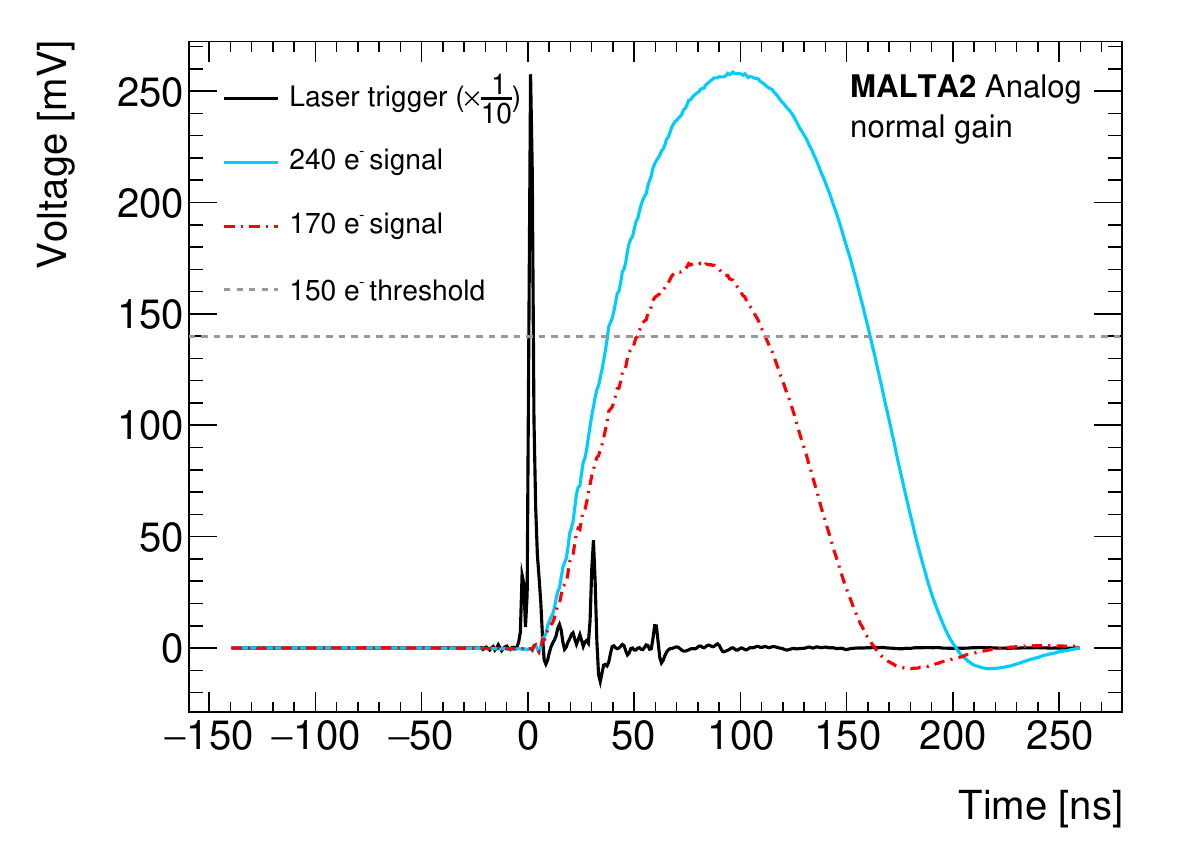}
\caption{Analog waveforms.}\label{fig:TCT_waveforms}
\end{subfigure}
\begin{subfigure}[t]{0.49\textwidth}
\centering
\includegraphics[width=\linewidth]{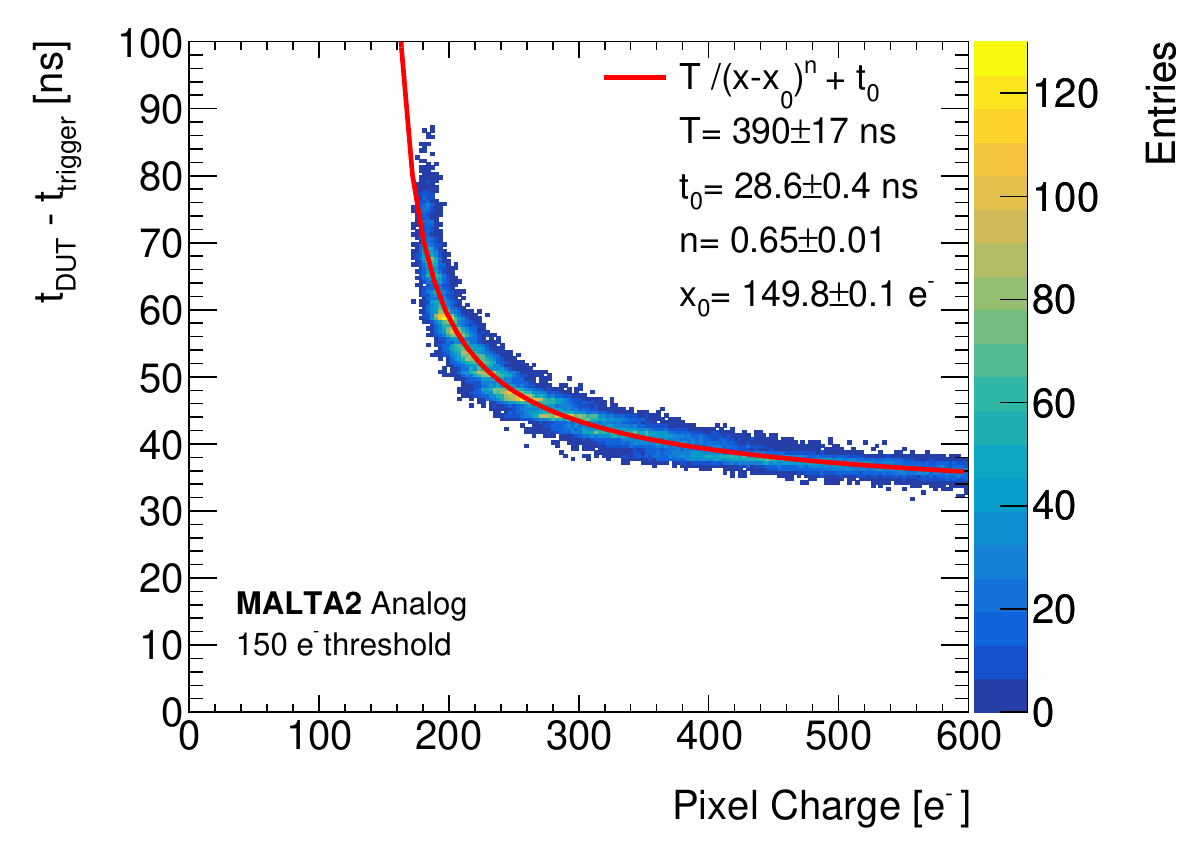}
\caption{Time walk curve at a threshold of $150\,\mathrm{e}^-$.}\label{fig:TCT_timewalk}
\end{subfigure}
\caption{Analog waveforms (a) are acquired with dedicated analog monitoring pixels of MALTA2 for different laser intensities with an E-TCT setup.
The time walk relative to the laser trigger is shown in (b).}\label{fig:TCT_waveforms_timewalk}
\end{figure*}

Pixel sensors for tracking that apply a fixed detection threshold are subjected to the time walk effect, thereby introducing a charge-dependent timing shift.
In order to obtain a parametric description as simulation input, the voltage amplitude of a MALTA2 pixel is measured with an oscilloscope for different intensities of an infrared laser.
Two characteristic waveforms  are shown in Figure \ref{fig:TCT_waveforms} for signal amplitudes of 170 and $240\,\mathrm{e}^-$.
The oscilloscope is triggered on the rising edge of the laser trigger at a voltage of \SI{1.4}{\volt}, which serves as a time reference.
The time of arrival of the device under test ($\mathrm{t_{DUT}}$) is defined as the time when the waveform crosses a fixed threshold.
The threshold voltage of \SI{140}{\milli\volt} is chosen in order to match the nominal calibrated value of $150\,\mathrm{e}^-$.
The time walk curve shown in Figure \ref{fig:TCT_timewalk} is obtained by focusing the laser on the pixel center, varying its intensity and measuring $\mathrm{t_{DUT}}$ relative to the laser trigger for each injected charge.
The fit function diverges when the pixel charge approaches the threshold. 
For large signals, the time walk curve asymptotically approaches a constant value of \SI{29}{\nano\second}. 
The typical charge generated by a minimum-ionizing particle in \SI{30}{\micro\meter} of silicon is $\sim1800\,\mathrm{e}^-$.
This shows that charge depositions close to a pixel center that are not shared among other pixels are less affected by the time walk. 
The time walk mainly shifts in time small charge depositions that arise due to charge sharing.
As a consequence, pixel clusters show a leading timing from the seed pixel followed by hits from neighboring pixels at later times.

As the time walk curve depends on the threshold value, the measured curve at a threshold of $150\,\mathrm{e}^-$ of Figure~\ref{fig:TCT_timewalk} serves as the basis for extrapolation.
The time walk $\Delta t$ for a charge $x$ can be estimated by scaling to any threshold value according to
\begin{equation}
\label{eq:timewalk_scaling}
\Delta t(x;\,\mathrm{thr}) = \frac{\SI{390.}{\nano\second}}{(x\,\frac{150\,\mathrm{e}^-}{\mathrm{thr}}-149.8)^{0.65}}+\SI{28.6}{\nano\second}. 
\end{equation}
The scaling is justified due to the MALTA2 front end design, in which a large threshold range can only be covered by changing the gain of the front end.
In this way, the amplification decreases with threshold.

There is an additional time delay in the digital read-out that depends on the pixel position.
A maximum of \SI{7}{\nano\second} accumulates when passing a hit along all 512 rows of a MALTA sensor. 
The corresponding delay per row is \SI{0.0137}{\nano\second}~\cite{MilouRAD}.

\subsection{Digital hit merging}\label{Sec:2.4}

\begin{figure*} 
\centering
\includegraphics[width=\linewidth]{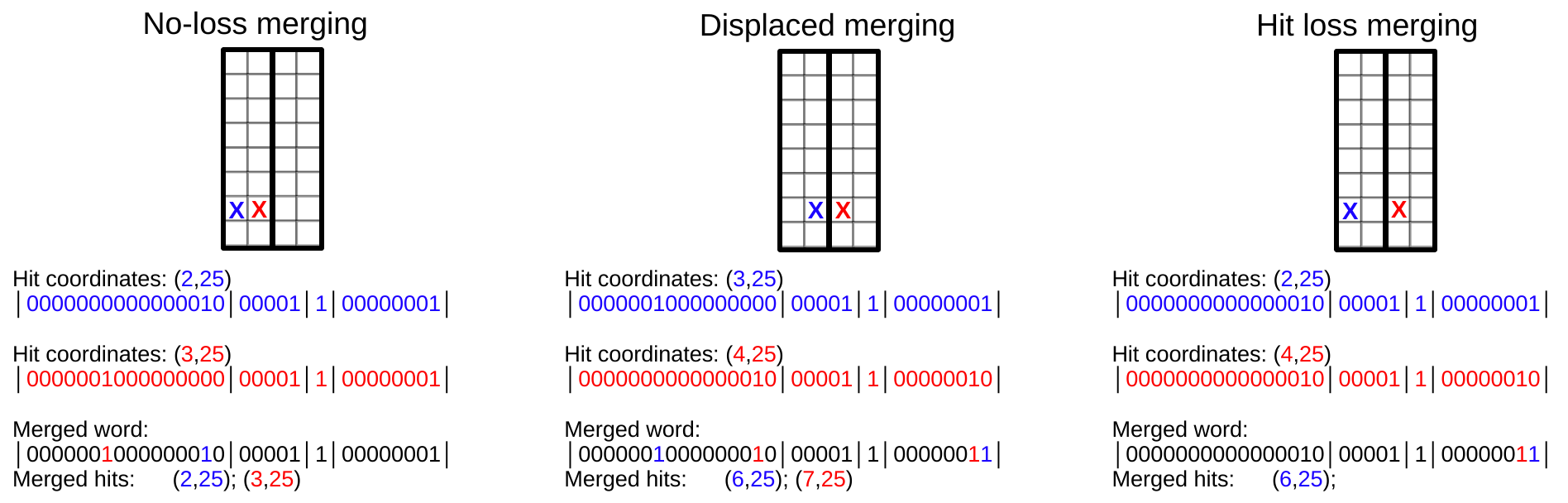}
\caption{Examples of merging of hits arriving within \SI{1.6}{\nano\second}. 
Each hit is associated with a 40 bit word that encodes its position in the in-pixel circuitry.
In the illustration, a reduced 30 bit word is shown, associated to the minimum amount of bits needed to encode a pixel position in the MALTA2 matrix.
For clarity, each bit group is delimited by vertical bars. 
The $16$ most significant bits hot-encode the pixel address.
The next $5$ bits encode the group address and the next $1$ bit the group parity.
The $8$ least significant bits encode the double column address.
For simplicity, the bits representing the reference, delay count, bunch crossing ID and chip ID are omitted here.
In the first case (left) the two hits are detected in the same $2\times8$ pixel group and are merged correctly because their encoding only differs in the pixel address. 
A cluster of two hits that spreads over different pixel groups (middle) is merged to two displaced hits.
Any two hits with the same pixel address (right) are merged to form a single hit which can be displaced.}\label{fig:MergingExamples}

\end{figure*}

\begin{figure*}
\centering
\begin{subfigure}[t]{\textwidth}
\centering
\includegraphics[width=0.6\linewidth]{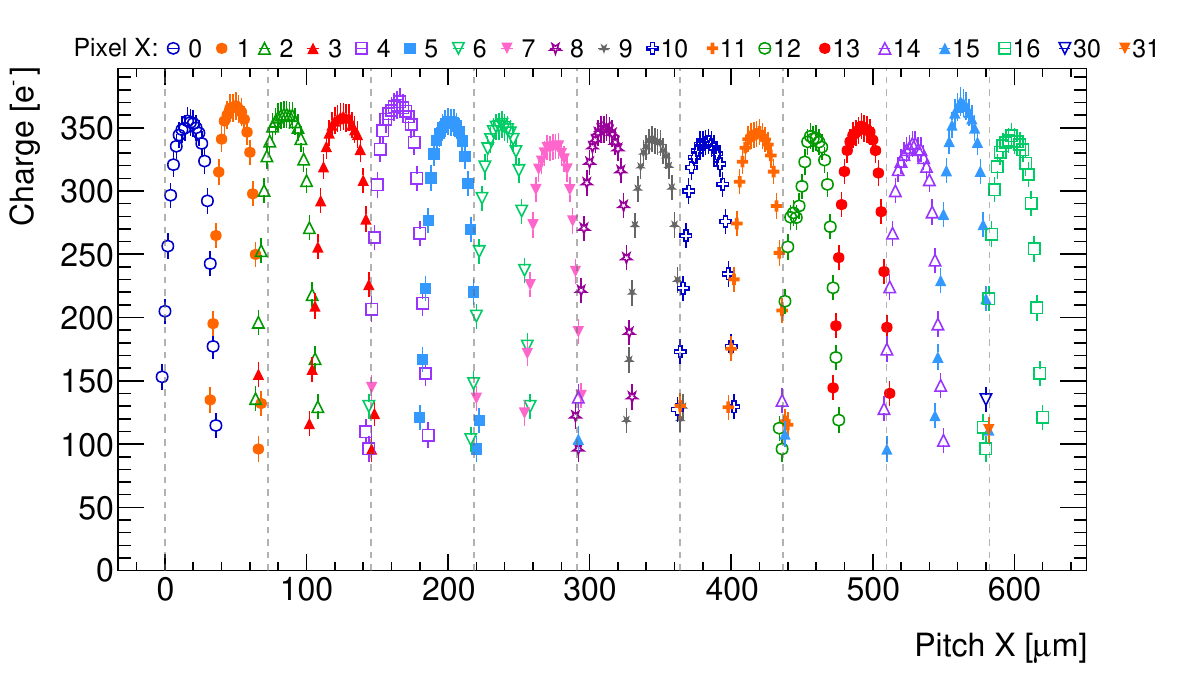}
\caption{Results of a pitch axis scan across pixels from column 0 to 16}\label{fig:LongPitchScan}
\end{subfigure}
\begin{subfigure}[t]{\textwidth}
\centering
\includegraphics[width=0.6\linewidth]{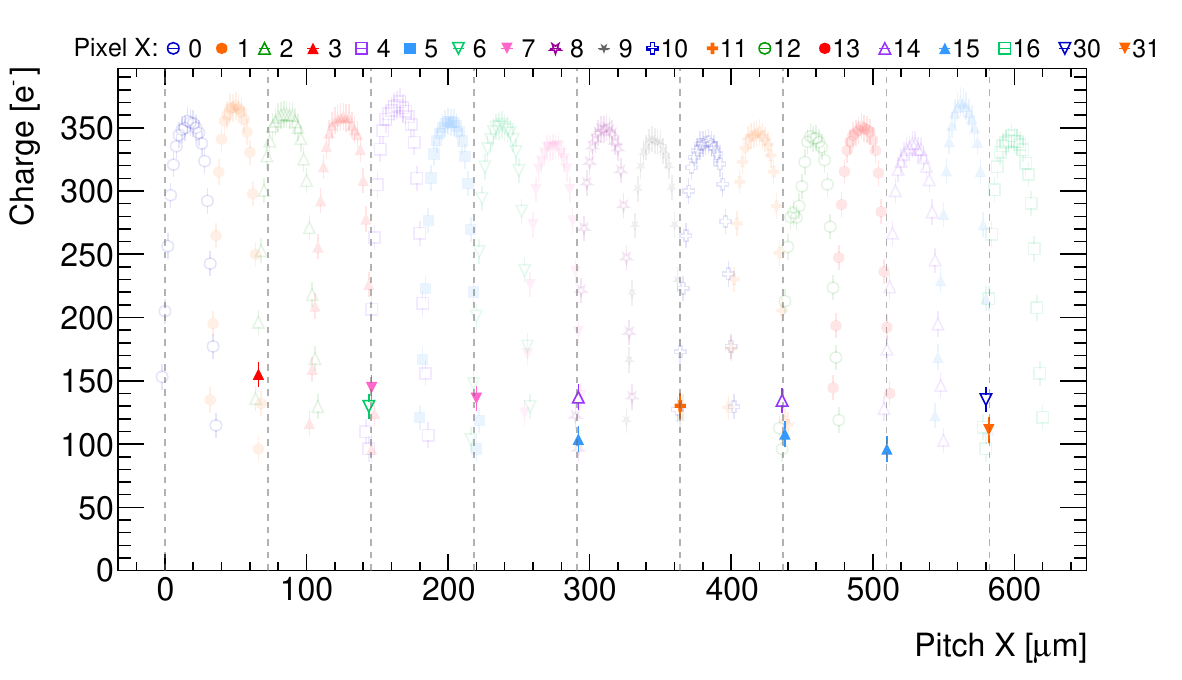}
\caption{Highlighted results of a pitch axis scan of hits appearing due to merging.}\label{fig:LongPitchScan_MergingHighlighted}
\end{subfigure}
\caption{Reconstructed charge of pixels within one row of a MALTA2 sensor measured with an E-TCT setup while scanning the pitch axis. 
(a) shows all reconstructed pixel charges. 
(b) highlights those data points that arise at least one pixel pitch displaced from the expected pixel center.
Dashed lines indicate the boundary of double-columns.
It is observed that displaced charge values appear at these boundaries due to hit merging.
The appearance of data from columns 30 and 31 is only due to merging.
Table~\ref{tab:HitMerging} summarizes all merged hits.}\label{fig:TCTPitchScanMerging}
\end{figure*}

The implementation of the MALTA2 asynchronous read-out requires a sub-division of the pixel matrix into double columns, which are further divided in $2\times8$ pixel groups. 
Each double column features two independent column drain buses that connect the pixel groups to the sensor periphery, dependent on their parity.
A hit above threshold triggers a 40 bit MALTA word. Its composition is listed in Table~\ref{tab:MALTAWord}.
Hits within the same group can be read out in one MALTA word. 
The corresponding pixel position is encoded by a 1 in the pixel address. The group and double column positions are encoded using 14 additional bits.
All words arriving within the same \SI{1.6}{\nano\second} time window are merged by a bitwise OR gate.
Figure~\ref{fig:MergingExamples} illustrates three cases of hit merging: no-loss merging, displaced merging and hit loss merging.
Experimentally, the merging can be studied through E-TCT by inducing a signal at the boundary of two pixel groups.
A one dimensional scan across the pitch axis is shown in Figure~\ref{fig:LongPitchScan} covering pixels from column 0 to 16.
Due to a slight inter-pixel gain variation in the analog front end, the maximum charge varies by an order of 10\% between pixels.
Figure~\ref{fig:LongPitchScan} also shows that a similar charge distribution is retrieved for each pixel with values of the FWHM compatible with the pixel pitch (\SI{36.4}{\micro\meter}).
Dashed lines mark the boundaries between pixel groups that are defined by the double-column arrangement.
Figure~\ref{fig:LongPitchScan_MergingHighlighted} shows the same dataset and highlights the data points that fall at least one pixel pitch away from the expected pixel center.
All of these points are associated to displaced merging occurring at the boundary between pixel groups.
In this case, the dominant merging effect is in the 8 bit long double column ID.
Table~\ref{tab:HitMerging} provides a visual aid to the displaced merging observed in the E-TCT data. 
It summarizes the column coordinate of two neighboring pixels and their double column ID in binary representation.
The same double column ID is shared by two neighboring columns in the same pixel group defined by the integer division
\begin{equation}
    \mathrm{dcol} = \mathrm{column}/2 .
\end{equation}
The merging between neighboring groups has a systematic pattern. 
All the displaced hits marked with bold in Table~\ref{tab:HitMerging} are observed in the highlighted data of Figure~\ref{fig:LongPitchScan_MergingHighlighted}. 
Every second group boundary leads to a single displaced hit (e.g. columns 1,2 $\rightarrow$ 2,3) where one hit remains unchanged.
Consequently, this surviving hit is still efficiently detected.
Every other boundary leads to two displaced hits (e.g. columns 3,4 $\rightarrow$ 6,7).
Such merging introduces detection inefficiencies if the displacement of both hits is larger than a specified tracking cut.\\

The merging has been tested for the specific case of scanning across the column axis of a MALTA2 sensor.
The appearance of displaced hits is explained by the digital hit merging which can introduce detection inefficiencies.
The effect of the merger onto the sensor efficiency will be discussed in Section~\ref{sec:TrackEffValid} and on the cluster size in Section~\ref{sec:ClSizeValid}.

\begin{table}
\centering 
\caption[The MALTA word]{The MALTA word composition of bits encoding a hit.}
\begin{tabular}{rrl}
    Bits & Size & Description   \\
    \hline
    0     & 1 &   Reference  \\
    1-16  & 16  &  Pixel address in $2\times 8$ pixel group\\
    17-21 & 5  &  Group address     \\
    22    & 1  &  Parity  (0 or 1)   \\
    23-25 & 3  &  Delay count     \\
    26-33 & 8  &  Double column ID (up to 256) \\
    34-35 & 2  &  Bunch crossing ID    \\
    36-39 & 4  &  Chip ID     \\
   \end{tabular}\label{tab:MALTAWord}
\end{table}

\begin{table}
\centering 
\caption{Hit merging across different double columns. 
Two pixel hits from neighboring columns are merged to output hits at different pixel coordinates.
The 4 least significant bits of the 8-bit double column IDs of both hits are listed. 
The upper 4 bits are zero and thus omitted. 
A bitwise OR operation is performed in the merger.
From the resulting double column ID, the corresponding merged column coordinates are obtained.
These columns always lie next to each other inside the same double column as they share the same MALTA word.
In bold are marked the output columns corresponding to displaced hits that are observed in Figure~\ref{fig:TCTPitchScanMerging}.}
\begin{tabular}{lcr}
    Hit column & Double column ID & Merged column\\
    input & OR operation & output\\
    \hline
    \,\,\,1, \,\,2 & 0000 $\vert$ 0001	$\rightarrow$ 0001 & \,\,2, \,\,\,\textbf{3}\\
    \,\,\,3, \,\,4 & 0001 $\vert$ 0010	$\rightarrow$ 0011 & \,\,\textbf{6}, \,\,\,\textbf{7}\\
    \,\,\,5, \,\,6 & 0010 $\vert$ 0011	$\rightarrow$ 0011 & \,\,6, \,\,\,\textbf{7}\\
    \,\,\,7, \,\,8 & 0011 $\vert$ 0100	$\rightarrow$ 0111 & \textbf{14}, \textbf{15}\\
    \,\,\,9, 10 & 0100 $\vert$ 0101	$\rightarrow$ 0101 & 10, \textbf{11}\\
    11, 12 & 0101 $\vert$ 0110	$\rightarrow$ 0111 & \textbf{14}, \textbf{15}\\
    13, 14 & 0110 $\vert$ 0111	$\rightarrow$ 0111 & 14, \textbf{15}\\
    15, 16 & 0111 $\vert$ 1000	$\rightarrow$ 1111 & \textbf{30}, \textbf{31}\\
   \end{tabular}\label{tab:HitMerging}
\end{table}

\section{The Simulation Approach}

The simulation package is structured in two distinct parts: the GEANT4 online simulation and the C++ offline analysis.
The GEANT4 simulation design principles revolve around a scalable and time efficient simulation (\SI{10}{\kilo Events/s}), while most computationally intensive tasks are relegated to the offline analysis.
This allows for an optimized work flow for investigating different sensor parameters and allows for efficient device scaling for large detector simulations.

\subsection{GEANT4 simulation}

The GEANT4 simulation implements a MAPS sensor as a monolithic silicon block with the dimension of the sensor's sensitive volume.
The silicon pixelation is applied for each GEANT4 hit together with the charge sharing model presented in Section \ref{Sec:2.2} and the time corrections presented in Section \ref{Sec:2.3}.

The particle gun implementation assumes a monoenergetic beam population.
This assumption does not match exactly the experimental conditions of the SPS mixed hadron beam ($120 -\SI{180}{\giga\electronvolt}$ mixed hadrons).
The impact of the discrepancy on the data-to-simulation comparison is expected to be negligible.
This is because a similar, minimum-ionizing energy deposition is expected for hadrons in this energy range in thin silicon substrates (\SI{30}{\micro\meter}).
The particle gun implements a configurable particle frequency which allows for simulating the effect of particle pile-up.

In the following study, the simulation of a MAPS for tracking and for digital calorimetry has been performed.
Their implementation in GEANT4 is discussed below.

\subsubsection{Tracking simulation}
\label{sec:5.1}
The tracking simulation mainly aims to validate the chosen charge sharing model against experimental data.
In order to more accurately recreate the test beam conditions, six passive tracking planes have been implemented as bare silicon blocks, three before the device under test and three after.
Their positions were implemented faithfully to the real telescope arrangement in order to include the effect of beam scattering.

The particle gun aims to resemble the average particle content and energy recorded during the numerous test beam experiments with the MALTA telescope \cite{MALTATele} in the SPS beam line.
As such \SI{120}{\giga\eV/c} protons have been chosen as the primary particle.
In order to avoid position biases, the beam has a uniform distribution across the sensor.

\subsubsection{Digital calorimetry simulation}
\label{sec:5.2}
The simulation of MAPS for digital calorimetry is achieved by implementing a \SI{3.2}{\centi\meter} thick tungsten block in front of the sensor.
The tungsten thickness corresponds to \SI{9}{X_0}, that is the depth at which an electromagnetic shower induced by an electron of \SI{120}{\giga\eV} reaches its maximum.
This will then provide a worst case analysis of the saturation effect of the energy measurement because at this depth the hit density is at its maximum.

The primary particles are electrons in order to consider only the effects of the electromagnetic showers with varying monoenergetic distributions between \SI{5} - \SI{100}{\giga\eV}.
In order to limit the effect of the shower leakage, a narrow Gaussian electron beam of \SI{100}{\micro\meter} width centered on the sensor was chosen.

\subsection{Offline analysis}

The offline tracking analysis performed on the simulation output (position, time and energy) is functionally divided into four parts ($1$ - $4$): digitization, tracking, clustering and analysis.
The offline calorimetry analysis contains only $2$ steps: ($1$, $5$): digitization and calorimetry.
All these steps are further detailed below.

\begin{figure*}
\centering
\begin{subfigure}[t]{0.49\textwidth}
\centering
\includegraphics[width=\linewidth]{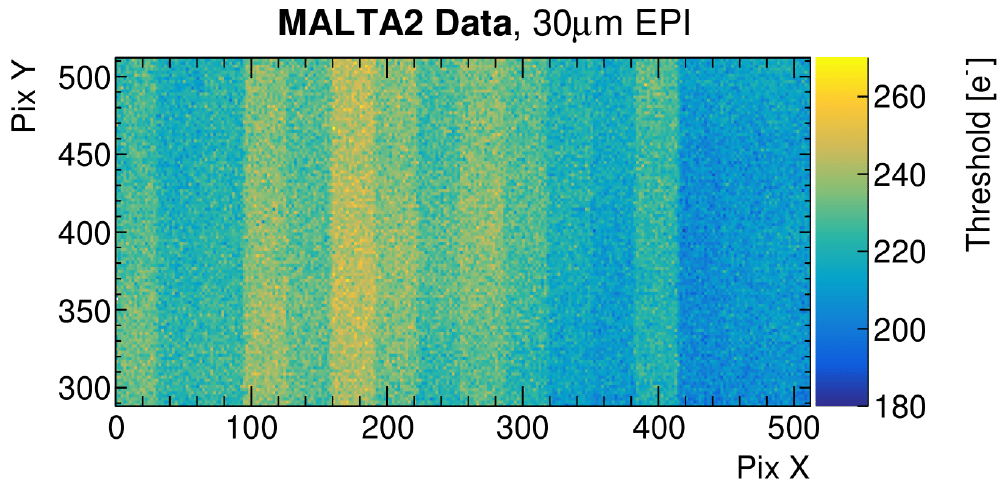}
\caption{Data.}
\label{fig:Threshold2D_Data}
\end{subfigure}
\begin{subfigure}[t]{0.4775\textwidth}
\centering
\includegraphics[trim=0 0 10 10, width=\linewidth]{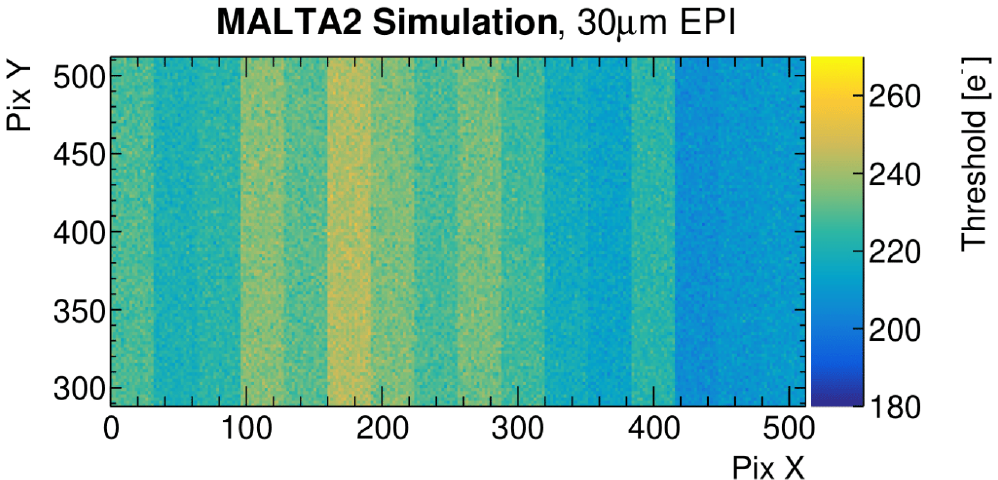}
\caption{Simulation.}
\label{fig:Threshold2D_Simu}
\end{subfigure}

\caption{Two-dimensional threshold distribution across a MALTA2 sensor.
The pattern arises due to the on-chip current mirror placement that insures a uniform propagation of the bias voltage across the sensor.}
\label{fig:Threshold2D}
\end{figure*}

\begin{figure}
\centering

\includegraphics[width=0.9\linewidth]{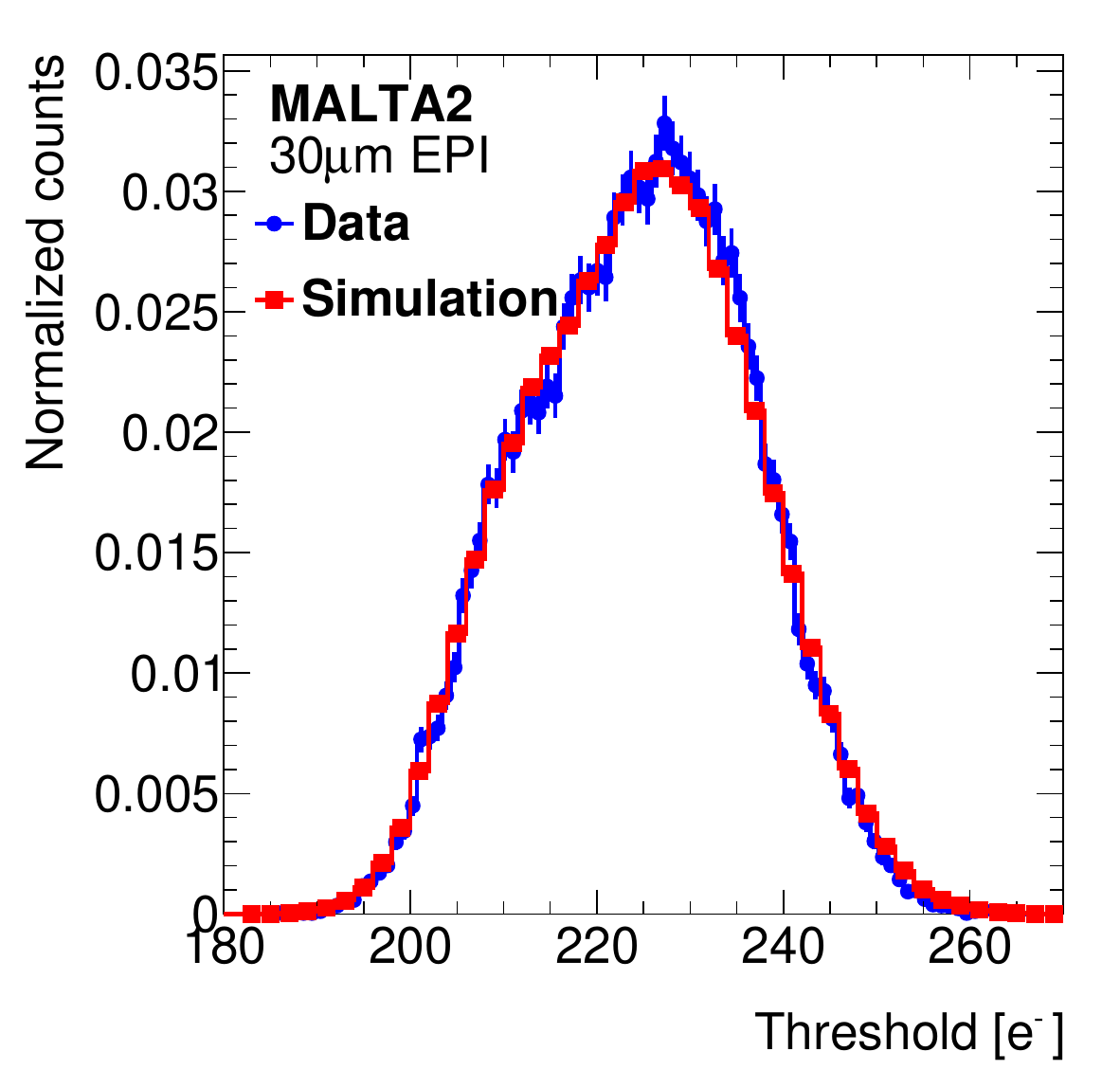}
\caption{Distribution of threshold values for data (blue line and points) and simulation (red line and points).}\label{fig:Threshold1D}

\end{figure}

\begin{enumerate}
    \item \textbf{Digitization.} Each GEANT4 entry is evaluated in terms of its energy deposition against the threshold of an individual pixel.
    A full chip threshold distribution for both data and simulation can be seen in Figure \ref{fig:Threshold2D}.
    The visible column pattern is well documented \cite{FPiroFE} and associated to the current mirrors present every $32$ columns in order to propagate the bias voltage across the sensor.
    The threshold dispersion varies chip-to-chip, but remains constant in time for one sensor.
    The simulation implements a random threshold distribution, however the mean for each $32$ column block was fixed for this particular data-to-simulation comparison.
    The threshold modeling is important in order to accurately describe the \SI{10}{\percent} dispersion in the threshold distribution observed in data \cite{BDVRadFE}.
    The match between data and simulation can also be observed in the threshold distribution of Figure \ref{fig:Threshold1D}.
    All charge depositions above the pixel threshold are considered as hits and are then encoded in terms of their pixel coordinates and time stamp according to the MALTA2 bit encoding.
    The merging logic described in Section \ref{Sec:2.4} is applied, followed by the decoding logic of the new pixel coordinates.

    \item \textbf{Tracking.} All decoded hits are matched to the primary particle tracked through the telescope set-up.
    The matching occurs in a fixed time window of \SI{500}{\nano\second} and spatial window of \SI{100}{\micro\meter} ($\sim3$ pixels).
    These values resemble the test beam analysis conditions.

    \item \textbf{Clustering.} Hits matched within the same time window undergo a clustering algorithm that checks for the formation of valid clusters, compatible with the definition in Reference \cite{Proteus}.
    Compatible to the test beam analysis, pixel clusters can only be formed from adjacent hits.
    Additionally, each cluster is endowed with a cluster timing, associated to the earliest hit.

    \item \textbf{Analysis.} Tracking figures of merit are computed: efficiency, cluster size and timing.

    \item \textbf{Calorimetry.} The decoded hit coordinates are matched to a primary particle if they fall in a time window defined by the beam frequency.
    The resulting number of secondary hits is then computed.
\end{enumerate} 

\subsection{Figures of merit}

The main figures of merit for tracking applications are the sensor efficiency, cluster size and timing.
The definition of these is compatible with previous definitions used in works describing the performance of the MALTA sensor \cite{MilouRAD}.
The average sensor efficiency is deduced as the ratio between hit-matched tracks and all tracks.
The hit timing is defined from the minimum cluster time relative to the mean timing. 
Additionally, the timing figure of merit is qualified in terms of the timing difference between a pixel's center and corner.

For the digital calorimetry application, the only figure of merit considered in this work is the number of detected secondaries (hits per event) which is approximately proportional to the energy of the primary scattered particle.

All plots include error bars given by the statistical uncertainty. 
Due to the large number of statistics in both data and simulation, these are often hidden by the marker size.

\section{Simulation validation}

The validation of the simulation has been performed with the help of complementary test beam data. 
The data was acquired with the MALTA telescope \cite{MALTATele} in the SPS beam line. 
A mixed beam of hadrons with energies around \SI{120}{\giga\eV} was used. 

The primary tracking figures of merit have been validated qualitatively and quantitatively in order to showcase the realism of the simulation. 

\subsection{Sensor efficiency validation}
\label{sec:TrackEffValid}

\begin{figure}[]
\centering

\includegraphics[width=0.95\linewidth]{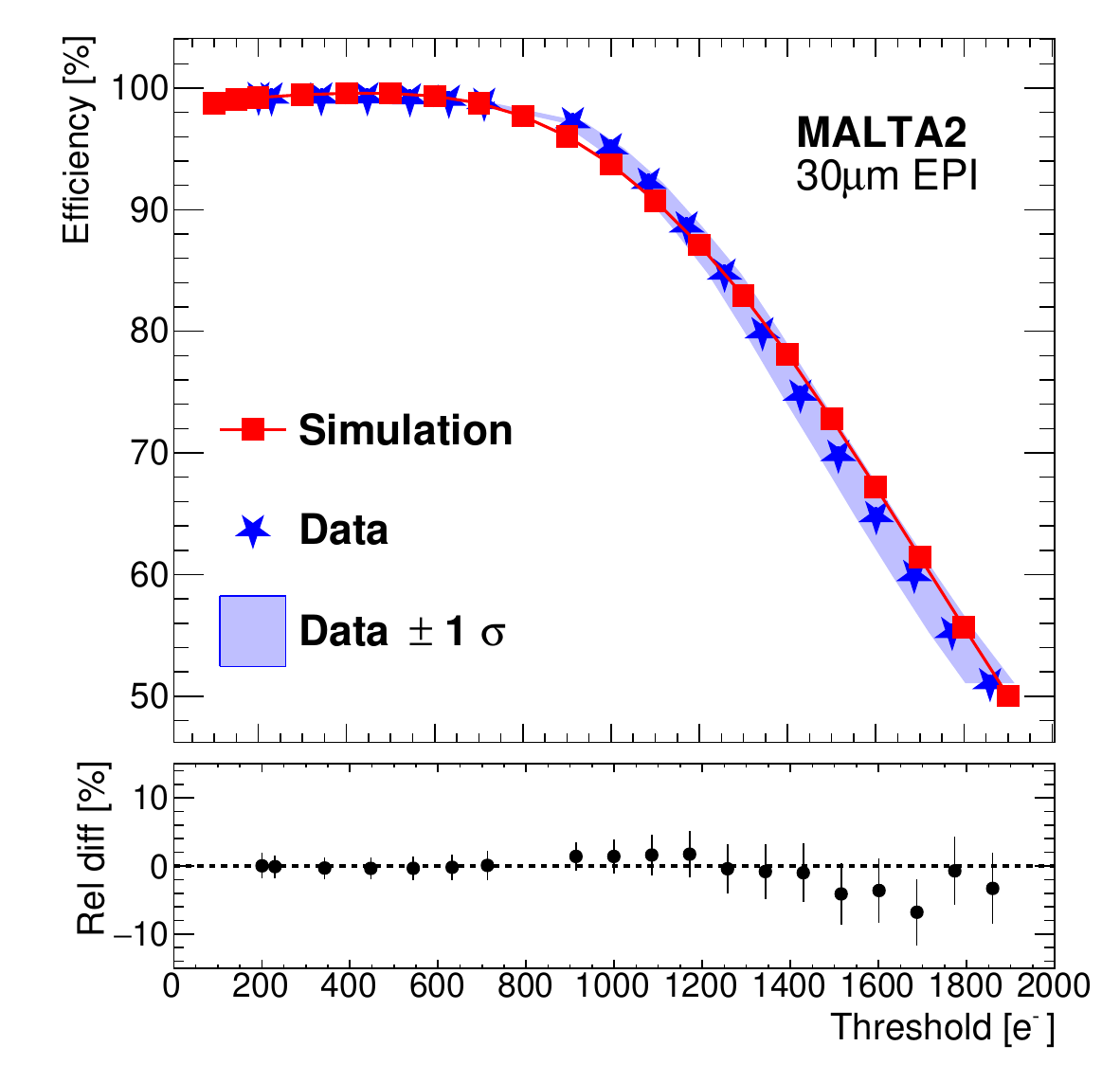}
\caption{Detection efficiency for various threshold sensor configurations. 
With blue, data from the SPS test beam is plotted together with a 1 sigma band dominated by the 3\% threshold uncertainty. 
With red, the simulation output is plotted. 
The relative data-simulation residual is plotted in the lower panel.}\label{fig:EfficiencyvsThreshold}

\end{figure}

Figure \ref{fig:EfficiencyvsThreshold} shows the efficiency for several threshold settings spanning the entire dynamic range of the sensor (\SI{200}{} - \SI{2000}{\mathrm{e}}$^-$). 
The data points have been obtained for the same sample, however in order to cover the large threshold range, two different front end configurations have been used, corresponding to a high and low gain configuration respectively \cite{FASSELTCHRECO}. 
A very good match between data (blue stars) and simulation (red squares and line interpolation) is obtained within a $1\sigma$ band, dominated by the \SI{3}{\percent} threshold uncertainty \cite{FASSELTCHCAL}. 
A relative difference in efficiency within $\pm$ \SI{2}{\percent} is gained for a sensor threshold below \SI{800}{\mathrm{e}}$^-$. 
There is a slight over-estimation of the efficiency in the simulation for sensor thresholds above \SI{1200}{\mathrm{e}}$^-$. 

\begin{figure*} 
\centering

\includegraphics[width=0.8\linewidth]{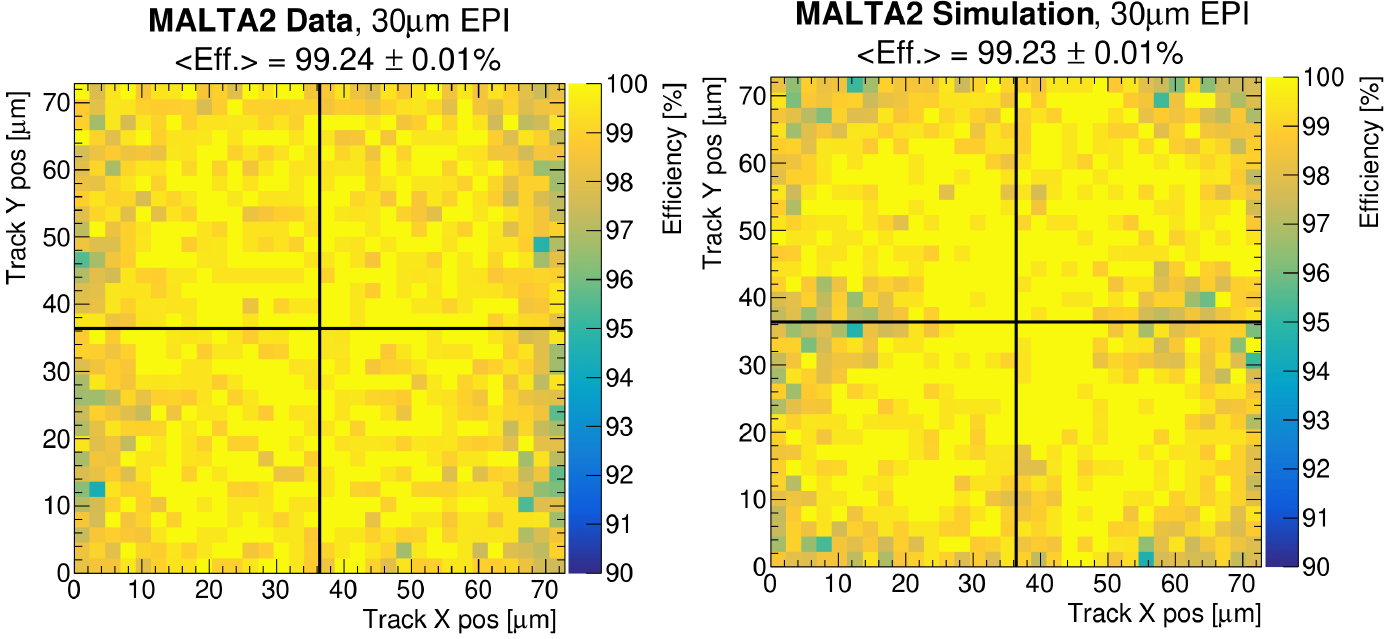}
\caption{Efficiency projected onto a $2\times2$ pixel matrix for data (left) and simulation (right). 
Both plots correspond  to a sensor threshold of \SI{200}{\mathrm{e}}$^-$. 
An average efficiency of $99.24\pm\SI{0.01}{\percent}$ was measured for data and $99.23\pm\SI{0.01}{\percent}$ for the simulation.}\label{fig:DataSim_2DEff}

\end{figure*}

Of particular interest is the nominal operational threshold of \SI{200}{\mathrm{e}}$^-$ applied to MALTA2 samples. 
For this particular sensor setting, an average efficiency of $99.24\pm\SI{0.01}{\percent}$ is obtained for data and $99.23\pm\SI{0.01}{\percent}$ for simulation.
While, the charge collection efficiency is expected to reach a value $>$ \SI{99.8}{\percent} at the pixel center, the lower efficiency measured and simulated is associated to the inefficiencies of the merging circuit. 
This highlights the accurate simulation of the merging circuit effect on the sensor's figures of merit. 
Figure \ref{fig:DataSim_2DEff} shows the efficiency for both data and simulation projected within a $2\times2$ pixel array. 
At the center of the pixel groups an efficiency $>$ \SI{99.98}{\percent} is gained for both simulation and data, matching the expectation of the technology node.
The inefficient in-pixel regions for both simulation and data are compatible with the displaced merging effect which occurs at the boundary of the $2\times8$ pixel groups but not within.

\begin{figure}[]
\centering

\includegraphics[width=0.9\linewidth]{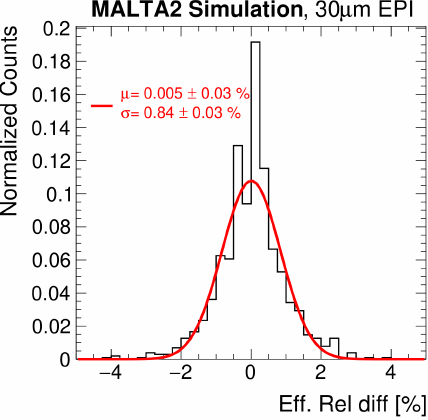}
\caption{The distribution of the relative efficiency residual between data and simulation.
A gaussian fit is plotted with red yielding  a mean of \SI{-0.005}{\percent} and an error on the mean of \SI{0.03}{\percent}.}\label{fig:Residual_1DEff}

\end{figure}

Figure \ref{fig:Residual_1DEff} showcases the one dimensional distribution of the data and simulation residual. 
A gaussian fit (red line) to the residual highlights a mean value of $-0.005\pm\SI{0.03}{\percent}$, which verifies the hit merger to equally effect data and simulation.

\subsection{Cluster size validation}
\label{sec:ClSizeValid}

\begin{figure}[]
\centering

\includegraphics[width=0.95\linewidth]{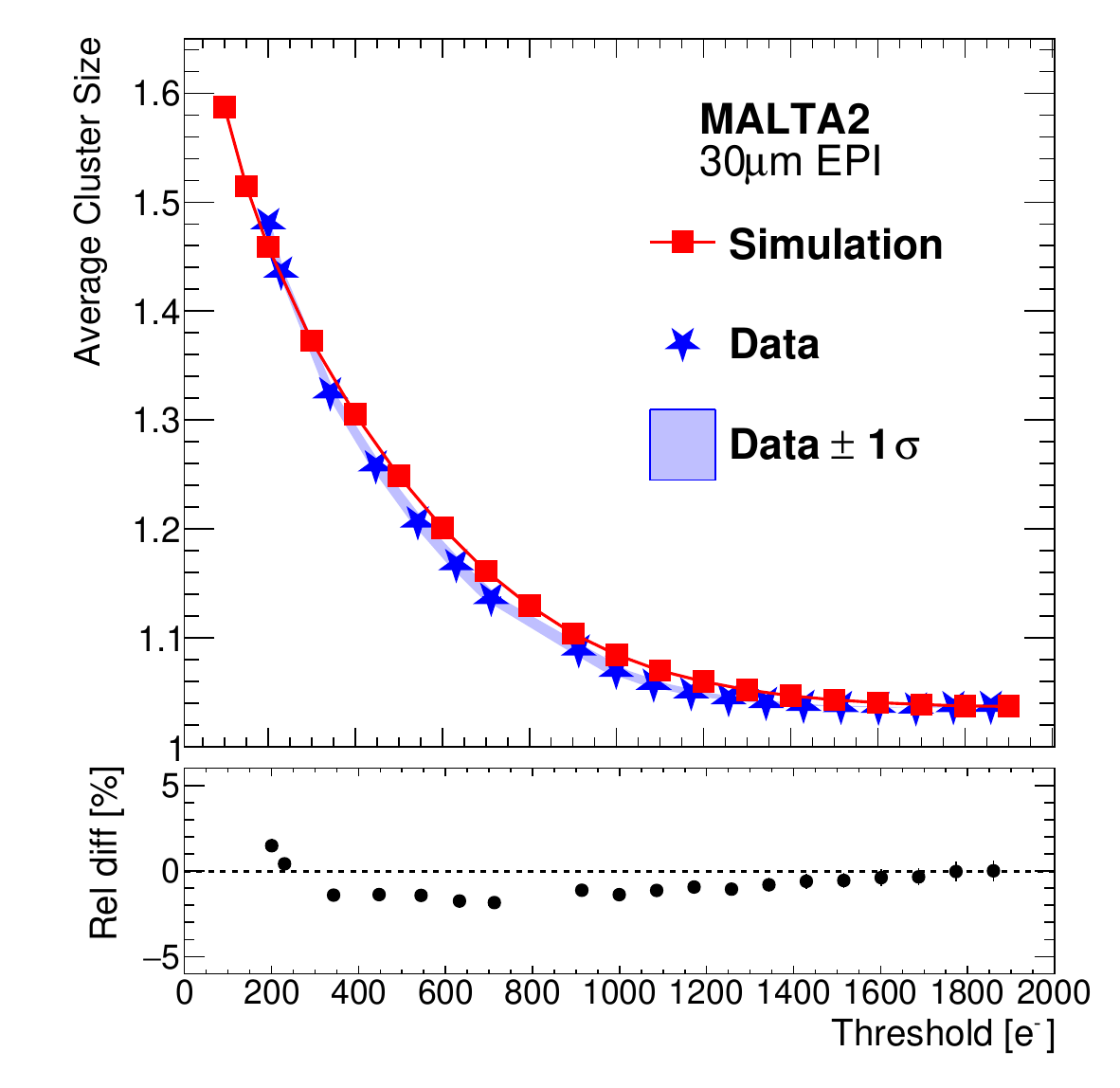}
\caption{Average cluster size for various threshold settings. 
With blue, data from the SPS test beam is plotted together with a 1 sigma band dominated by the 3\% threshold uncertainty. 
With red, the simulation output is plotted. 
The relative data-sim residual is plotted in the lower panel.}\label{fig:ClSizevsThreshold}

\end{figure}

A similar study has been performed for the average cluster size. 
Figure \ref{fig:ClSizevsThreshold} shows the average cluster size for several sensor threshold settings. 
A good match (within $1\,\sigma$) has been obtained between data (blue stars) and simulation (red squares and interpolated line). 
The computed residual, additionally showcases a data to simulation variation within $\pm$\SI{5}{\percent} across all threshold settings.

\begin{figure*} 
\centering

\includegraphics[width=0.8\linewidth]{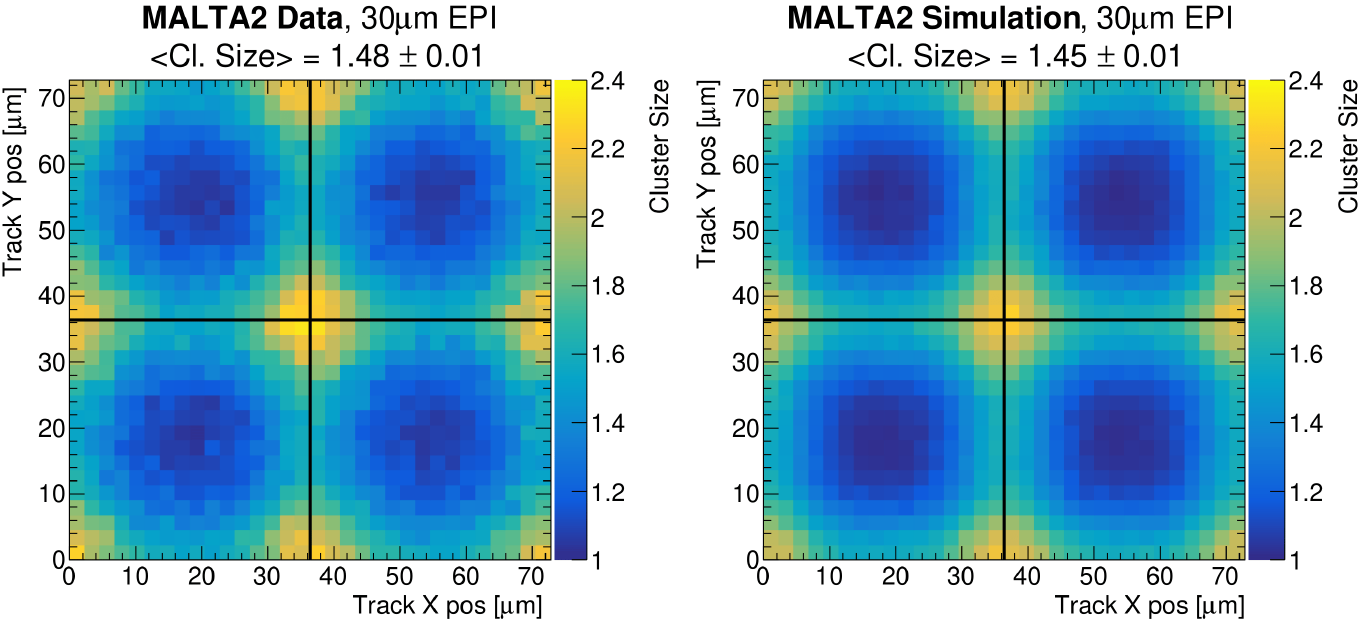}
\caption{Cluster Size projected onto a $2\times2$ pixel matrix for data (left) and simulation (right). 
Both plots correspond  to a sensor threshold of \SI{200}{\mathrm{e}}$^-$. 
An average cluster size of $1.48\pm\SI{0.01}{pixels}$ was measured for the data and $1.45\pm\SI{0.01}{pixels}$ for the simulation.}\label{fig:DataSim_2DClSize}

\end{figure*}

\begin{figure}[]
\centering

\includegraphics[width=\linewidth]{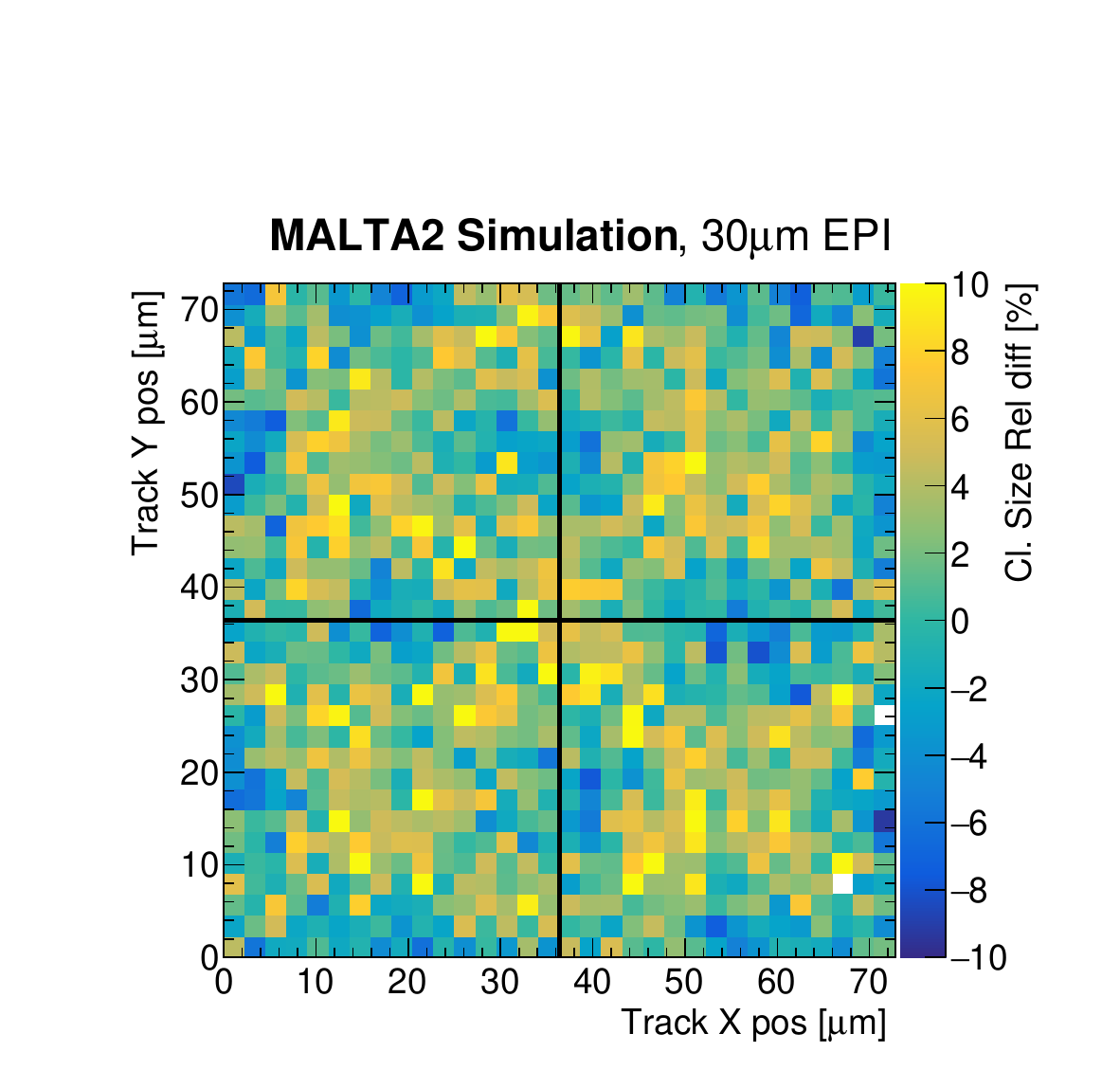}
\caption{Relative cluster size residual between data and simulation projected onto a $2\times2$ pixel matrix.}\label{fig:Residual_2DClSize}

\end{figure}

\begin{figure}[]
\centering

\includegraphics[width=0.95\linewidth]{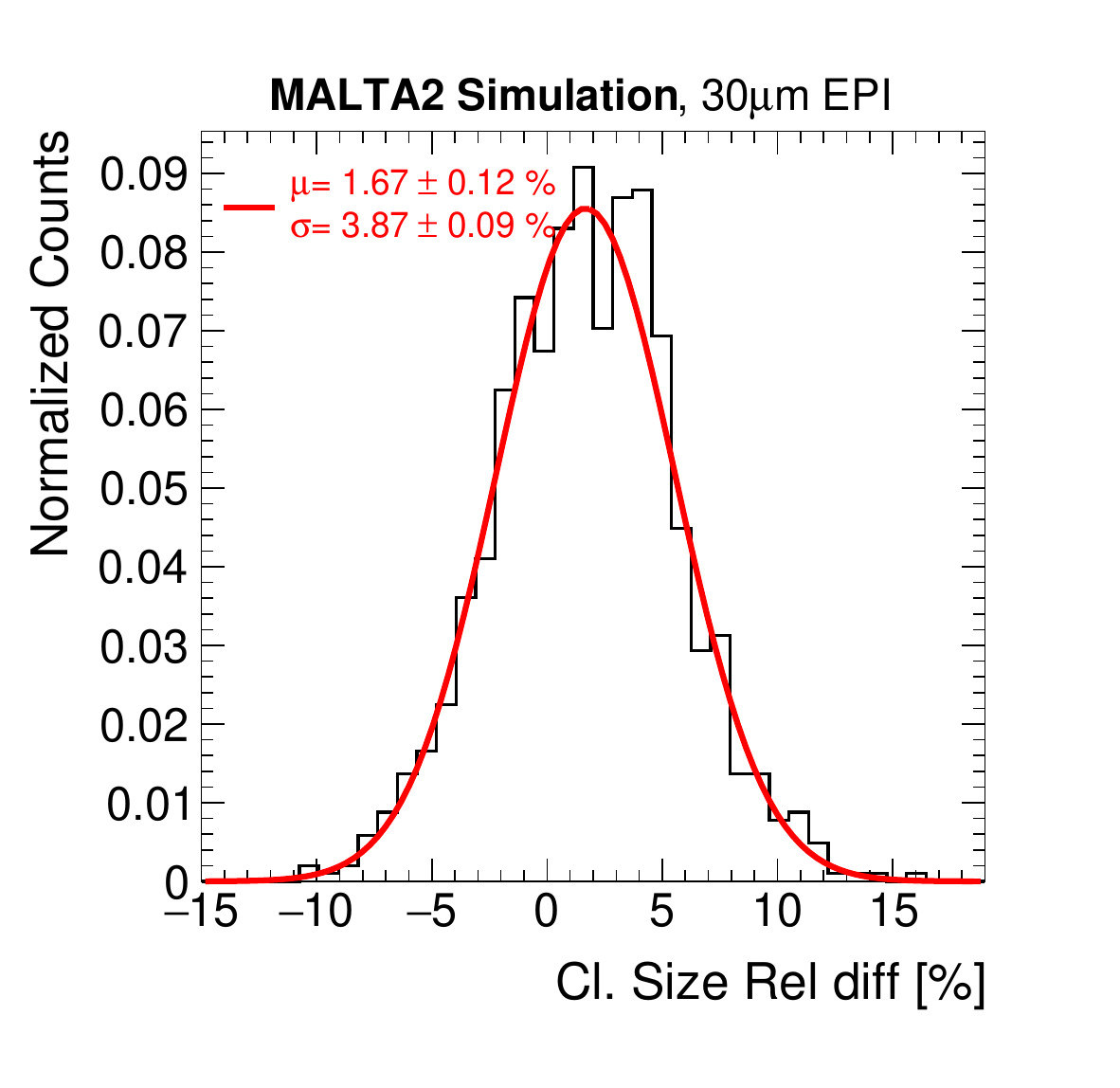}
\caption{The distribution of relative cluster size residual between data and simulation. 
A gaussian fit (red line) yields a mean of \SI{1.67}{\percent} and an error on the mean of \SI{0.12}{\percent}. }\label{fig:Residual_1DClSize}

\end{figure}

Considering the operational threshold of \SI{200}{\mathrm{e}}$^-$, the $2\times2$ pixel projections for data and simulation are shown in Figure \ref{fig:DataSim_2DClSize}. 
A quantitative match between the two is obtained: an average cluster size of $1.48\pm\SI{0.01}{pixels}$ is obtained for data, while the simulation records a value of $1.45\pm\SI{0.01}{pixels}$. 
The qualitative match between the two is apparent, however a more descriptive image of the residual distribution is obtained in Figure \ref{fig:Residual_2DClSize}. 
No significant pattern is observed, however a larger bin-to-bin variation is recorded compared to the efficiency residual in Figure~\ref{fig:Residual_1DEff}. 
A quantitative analysis of the residual is performed in Figure \ref{fig:Residual_1DClSize} where an average value of \SI{1.67\pm0.12}{\percent} was obtained. 
This highlights that the simulated cluster size is compatible with data within a deviation of \SI{2}{\percent}.

\subsection{Timing validation}

\begin{figure*}
\centering

\includegraphics[width=0.8\linewidth]{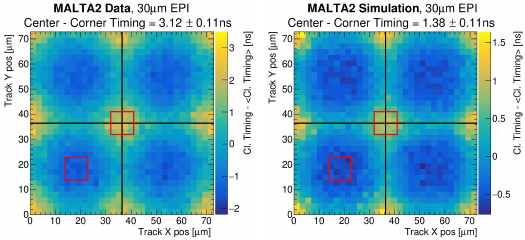}
\caption{Average subtracted cluster timing projected onto a $2\times2$ pixel matrix for data (left) and simulation (right). 
Both the data and simulation correspond to \SI{200}{\mathrm{e}^-} threshold. 
The center to corner timing difference was computed using the bins indicated by the 2 red squares. 
A value of \SI{3.12\pm0.11}{\nano\second} was obtained for the data and \SI{1.4\pm0.11}{\nano\second} for the simulation.}\label{fig:DataSim_2DTiming}

\end{figure*}

Figure \ref{fig:DataSim_2DTiming} shows the $2\times2$ in-pixel timing distributions for data and simulation. 
The red boxes denote the bins considered in the computation of the average center and corner timing respectively. 
A center to corner timing difference of $3.12\pm\SI{0.11}{\nano\second}$ is obtained for data, while a value of \SI{1.4\pm0.11}{\nano\second} is obtained for the simulation. 
This highlights that the two do not have an exact quantitative match, motivated by the simplifications made in the timing model. 
However, a very good qualitative match between simulation and data is obtained. 
The differences between simulation and data are explained by the current timing and threshold scaling model described in Section \ref{Sec:2.3}, which assume that a threshold change is accompanied by a front end gain change.

\section{Digital front end simulation testbed}

The MALTA2 sensor is expecting a redesign in the \SI{65}{\nano\meter} feature size TPSCo technology \cite{TPSCo}, in order to benefit from the increased intrinsic radiation hardness and front end active element density. 
A key part of the redesign is addressing the main inefficiencies of the previous sensor iterations and bringing  improvements to the design of the digital read-out. 
In this study we performed an investigation of different digital front end parameters in order to gauge the current design limitations and possible improvements for future MAPS. 
Only a conservative variation of sensor digital parameters was considered, in order to maintain a realistic sensor design.

One of the novel features of the MALTA sensor is the asynchronous read-out. 
A detector employing asynchronous sensor modules offers the prospect of a trigger-less and low material budget detection system (due to the limited supporting infrastructure). 
An unintended consequence of an asynchronous read-out is the requirement of prioritizing or gracefully merging multiple coincidental hits at the sensor's periphery. 
As was already shown in the previous sections, the multiple hit handling can introduce inefficiencies with different impacts depending on the sensor settings or beam conditions. 
With the advent of new and more demanding experimental requirements \cite{ECFA}, more MAPS design projects are considering the implementation of an asynchronous read-out. 
As a consequence, the simulation work provided in this section holds a significant value for the design of future sensors. 

Two different physics cases have been considered in the context of proposing new asynchronous read-out sensors: tracking and digital calorimetry. 
Each case is further discussed individually.

\subsection{Tracking application}

MAPS for tracking applications require sub pixel position resolution. 
As a consequence, large pixel coordinate shifts due to the merging circuit can introduce position resolution degradation. 
The same effect is expected to degrade the detection efficiency when hits are merged away by $3$ or more pixels ($>$ \SI{100}{ \micro\meter}). 

\begin{figure}[] 
\centering

\includegraphics[width=0.9\linewidth]{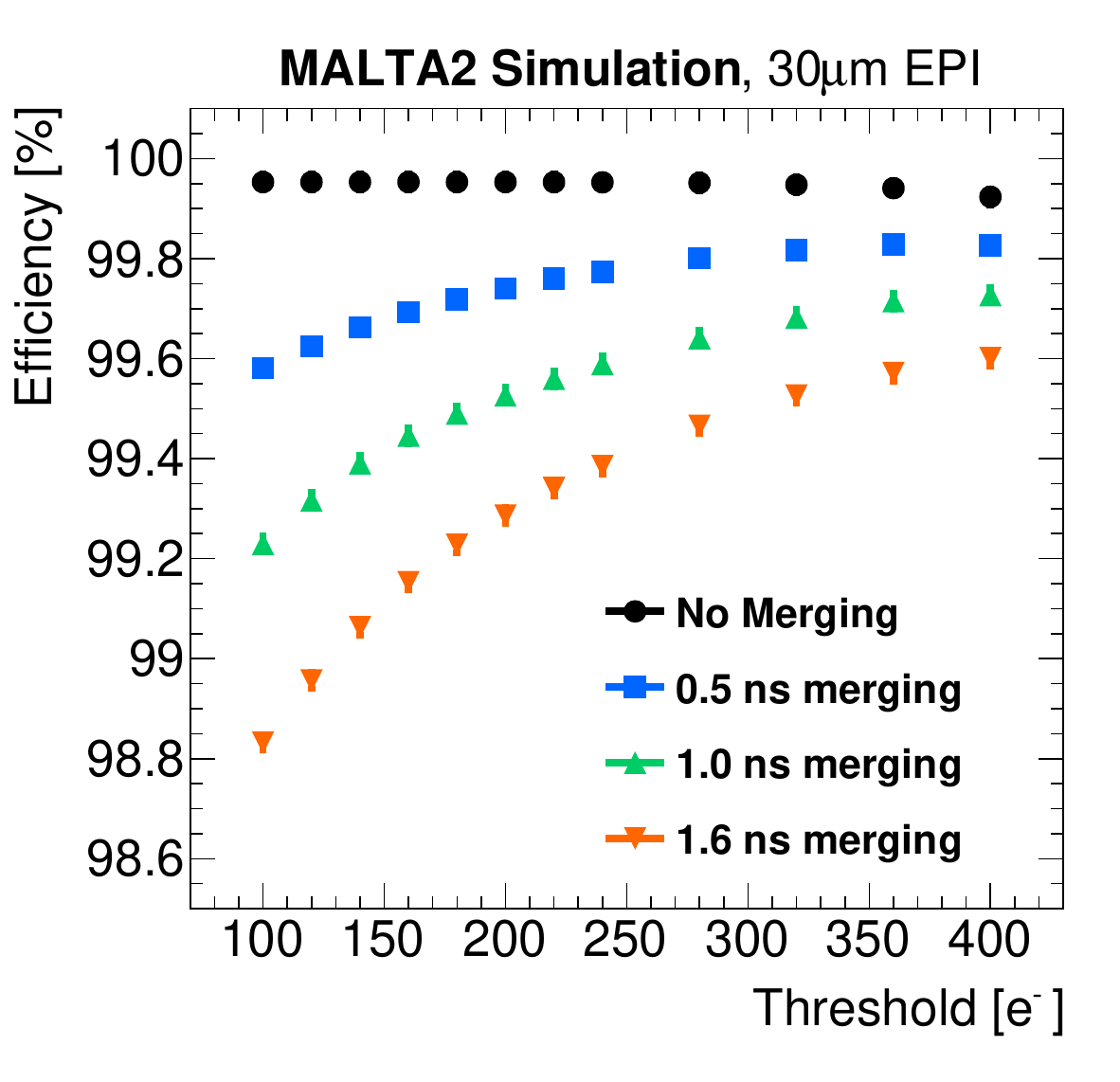}
\caption{Simulation of efficiency for various sensor threshold settings. 
The different curves correspond to various merging time windows. 
The time window corresponding to the no merging, functionally denotes the idealized case of a \SI{0}{\nano\second} merging window. 
All simulation points correspond to a $2\times8$ pixel grouping.}\label{fig:EffThr_MergingTime_UPDATED}

\end{figure}

One of the most important sensor parameters that is expected to impact the merging inefficiency is the merging window size. 
This value is inversely proportional to the clock frequency applied to the logic block.
Figure~\ref{fig:EffThr_MergingTime_UPDATED} shows the simulated efficiency for several thresholds within the expected operational range of a MALTA2 sensor (below \SI{400}{\mathrm{e}^-})~\cite{FASSELTCHCAL}. 
The several sets of points represent different merging window sizes ranging from the \SI{1.6}{\nano\second} (orange triangles) which matches the current sensor implementation, to the no merging implementation (black points) representing an ideal case where an infinite frequency clock is applied to the read-out. 
A slight efficiency decrease ($\sim$\SI{0.05}{\percent}) at thresholds above \SI{400}{\mathrm{e}^-} is caused by charge depositions at a pixel corner that get equally shared among its neighbors and thus fall below threshold. 
A systematic decrease in efficiency with an increase in the merging time window is explained by an increased time coincidence between time adjacent hits (shared charge or multiple primary particles). 
An increase in efficiency is gained with relatively higher thresholds up to a plateau ($\sim$\SI{400}{\mathrm{e}^-}), due to the inefficiencies being dominated by the merging of neighboring hits from charge sharing. 
Higher thresholds decrease the average cluster size as can be seen in Figure \ref{fig:ClSizevsThreshold}, leading to less frequent merging. 
Additionally, due to the interplay between timing and threshold, discussed in Section \ref{Sec:2.3} and highlighted in Equation \ref{eq:timewalk_scaling}, larger thresholds on average are associated to a larger time walk for spatially adjacent hits, leading to less frequent merging. 
A merging time window of \SI{0.5}{\nano\second} operated at a threshold of $\sim$\SI{300}{\mathrm{e}^-} leads to a significant efficiency improvement of \SI{99.8}{\percent}.

\begin{figure}
\centering
\includegraphics[width=\linewidth]{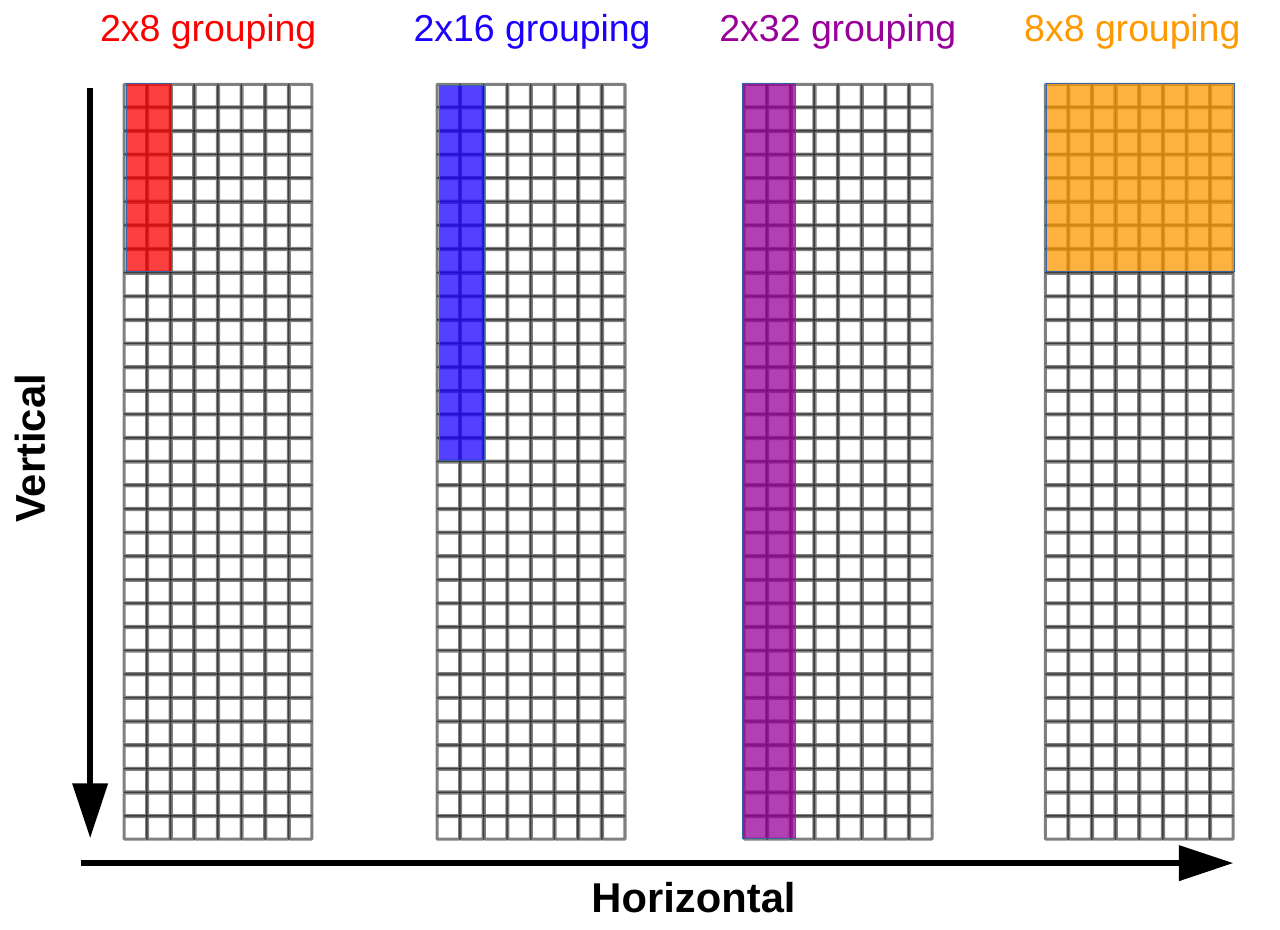}
\caption{Pixel grouping illustration. 
The $2\times8$ pixel group is the one implemented in the MALTA2 sensor.
Other groupings are studied in order to reduce the effect of the merging logic when merging hits from different groups.}\label{fig:Grouping}
\end{figure}

\begin{figure}[]
\centering

\includegraphics[width=0.9\linewidth]{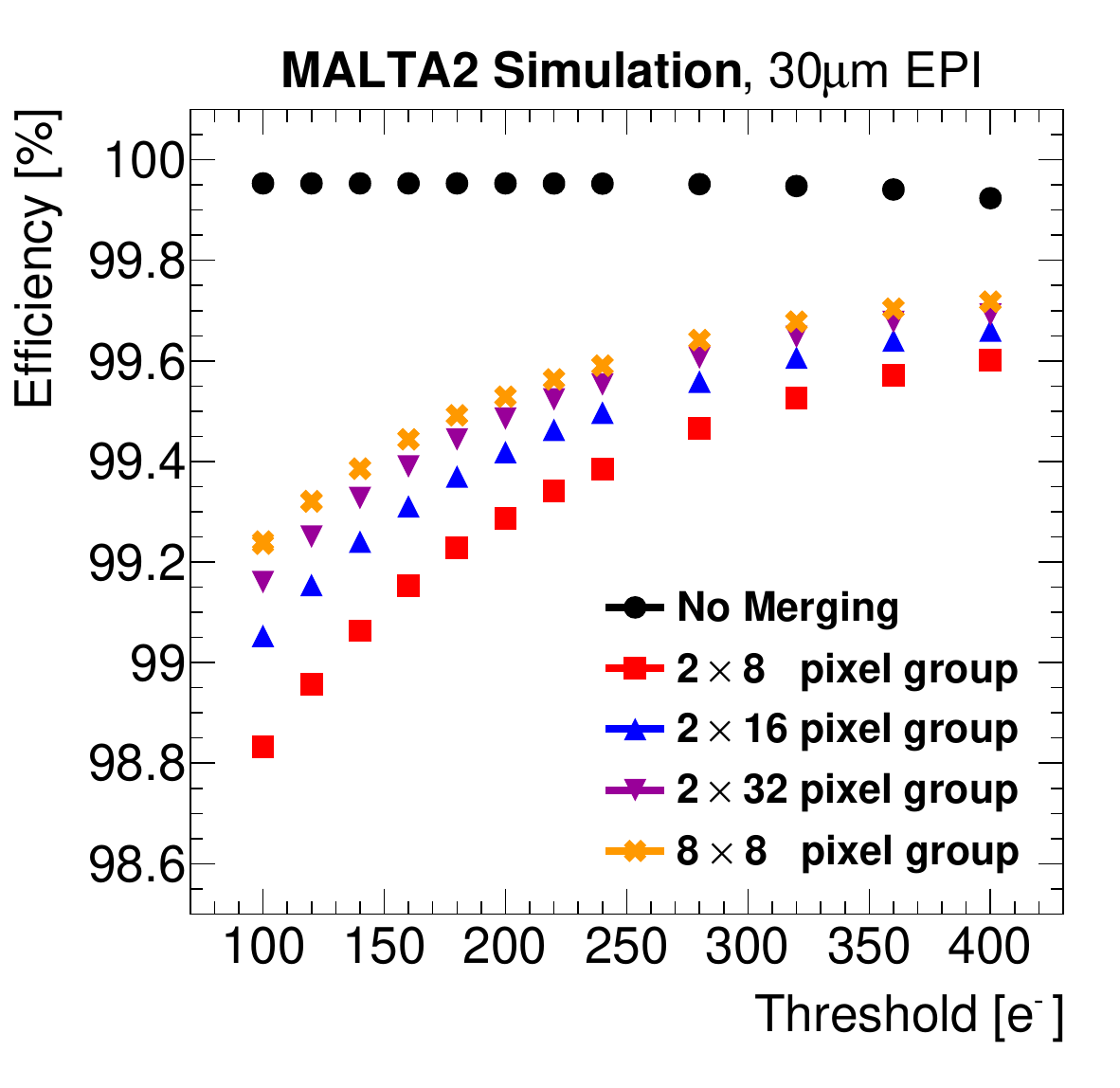}
\caption{Simulation of efficiency for various sensor threshold settings. 
The different curves are related to various pixel groupings in the digital read-out of the sensor. 
The $2\times8$ group size corresponds to the digital read-out in the real device, while other pixel groups correspond to possible sensor redesign choices. 
Additionally, the idealized case of no merging is plotted. 
All simulation points correspond to a \SI{1.6}{ns} merging window.}\label{fig:EffThr_GroupSize_UPDATED}

\end{figure}

Another strategy for mitigating the merging effect is by varying the pixel grouping in the read-out. 
Figure \ref{fig:EffThr_GroupSize_UPDATED} investigates the impact on the efficiency of pixel groupings of various sizes from the current design $2\times8$ pixel groups (red squares), to larger column groups of $2\times32$ (purple triangles) and also an $8\times8$ pad pixel group layout (orange crosses).
These groupings are illustrated in Figure~\ref{fig:Grouping}.
A larger pixel grouping is associated with an increase in efficiency. 
An additional effect can be observed for the $8\times8$ pixel grouping, where an increase in efficiency is gained over the $2\times32$ pixel grouping. 
Both highlighted effects can be understood from geometrical considerations. 
Hit merging happens only between neighboring pixel groups, scaling with the perimeter of the group. 
In contrast, no merging shifts occur for hits that occur in the same group. 
As expected, the $8\times8$ pad configuration presents the smallest perimeter among all investigated configurations of the same area, hence the highest efficiency. 
Overall the increase in the pixel group size shows to have a smaller impact on increasing the efficiency, compared to a higher frequency read-out clock. 
The most optimal pixel grouping for optimizing the efficiency was found to be the $8\times8$ pad grouping which is expected to reach an efficiency of \SI{99.6}{\percent}.

\begin{figure}[]
\centering

\includegraphics[width=0.9\linewidth]{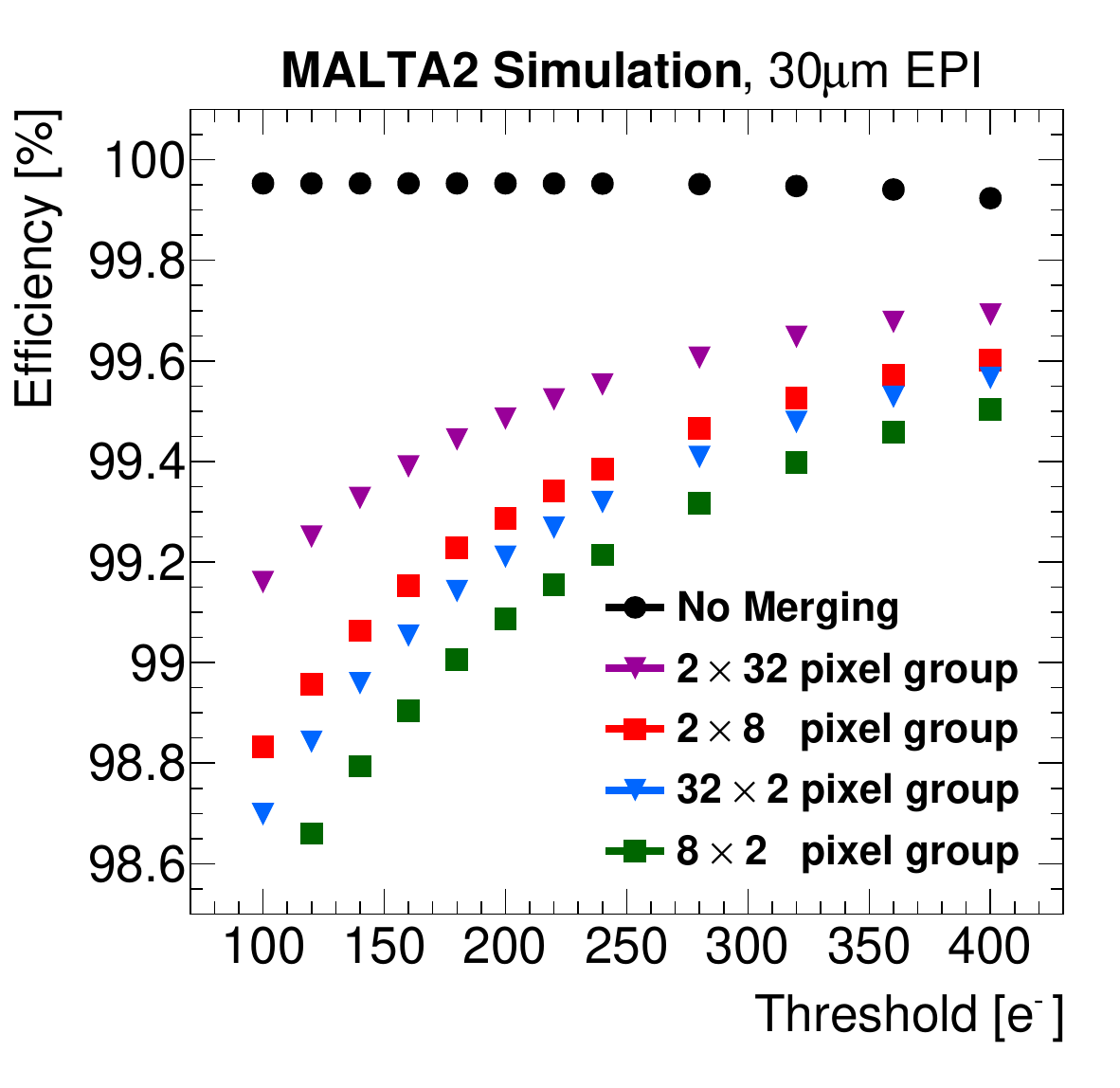}
\caption{Simulation of efficiency for various sensor threshold settings. 
The different curves correspond to various spatial group arrangements. 
A vertical pixel group arrangement is plotted for a $2\times8$ and $2\times32$. 
Additionally, the horizontal equivalent is plotted $8\times2$ and $32\times2$. 
The idealized case of no merging is also plotted. 
All simulation points correspond to a \SI{1.6}{ns} merging window.}\label{fig:EffThr_GroupParity_UPDATED}

\end{figure}

Figure \ref{fig:EffThr_GroupParity_UPDATED} shows the impact of two orientations of a $2\times8$ and $2\times32$ pixel groups: vertical (red squares and purple triangles) and horizontal (green squares and blue squares). 
A consistently higher efficiency is obtained for the vertical orientation of the pixel groups. 
This effect is associated to the asymmetric bit encoding between the column and row directions, discussed in Section \ref{Sec:2.4}. 
The additional parity bit that encodes the group position introduces a systematic shift proportional to the group Y dimension in merging shifts along this axis. 
In contrast, the column-wise merging introduces a variable bit shift between the group X dimension and up to 256 pixels. 
In the nominal case, this shift can be as small as $2$ pixels. Due to the \SI{100}{\micro\meter} tracking position cut, statistically fewer hits will be lost on this axis. 
This result shows the strength of the current read-out design. 
No improvement would be gained via a horizontal extension of the pixel groups, compared to a similar vertical extension.

\subsection{Digital calorimetry application}

Digital calorimetry promises a novel energy measurement technique for high energy particle physics \cite{PHALLDECAL}. 
The need for high granularity makes MAPS an obvious candidate for such an application. 
While digital calorimetry expects lower statistical uncertainties on the energy measurement, it also introduces unique challenges in terms of high rate particle counting. 
This challenge is magnified by the implementation of an asynchronous read-out. 
Several mitigation techniques are presented that aim to improve the energy reconstruction linearity.

Similarly to the tracking application, the main source of inefficiency in terms of secondary particle counting is the merging circuit. 
When only counting hits, the hit shifting does not negatively impact the study. 
In turn only hit loss merging is expected to impact the energy reconstruction.

\begin{figure}[]
\centering

\includegraphics[width=0.9\linewidth]{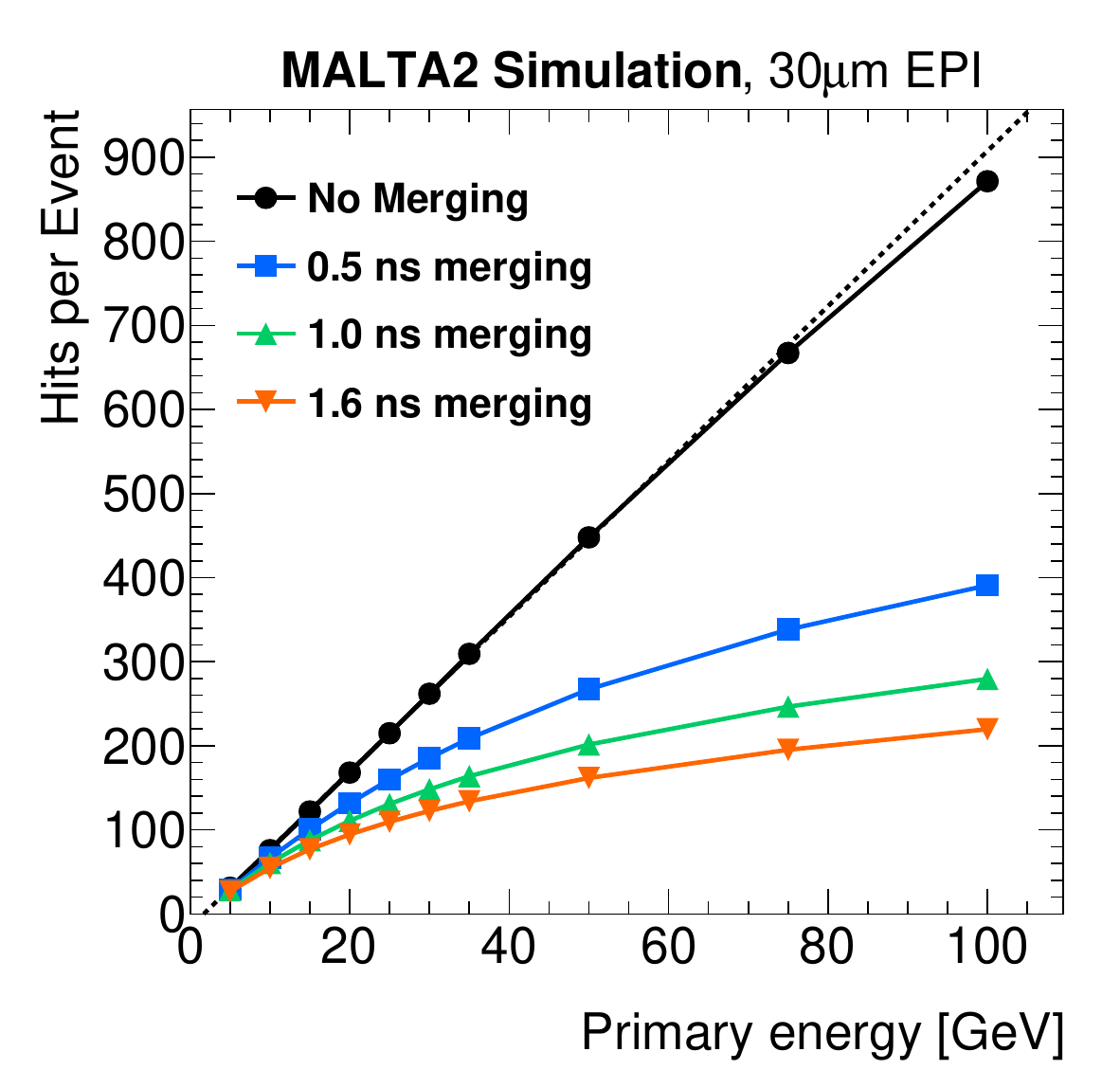}
\caption{Simulation of detected number of hits for an incoming electron of variable energy that interacts with a \SI{3.2}{cm} tungsten block. 
The different curves correspond to various merging time windows. 
The time window corresponding to the no merging, functionally denotes the idealized case of a \SI{0}{\nano\second} merging window. 
The black dotted line corresponds to a linear fit of the no merging data, performed for electron energies up to \SI{35}{GeV}. 
All simulation points correspond to a $2\times8$ pixel grouping.}\label{fig:CalovsWindowSize}

\end{figure}

Figure \ref{fig:CalovsWindowSize} shows the simulated mean number of detected hits for each primary particle interaction on a tungsten block. 
The black curve and points show the idealized case when no merging occurs in the sensor. 
It records a good linearity up to $\sim$\SI{80}{\giga\eV}. 
For higher energies, a slight saturation effect can be observed which is explained by the large hit multiplicity that can not be spatially resolved. 
The orange curve and points show the expected performance for the current sensor design (\SI{1.6}{ns} merging). 
It highlights that a large hit counting saturation occurs at relatively low energies ($<$\SI{20}{\giga\eV}). 
The decrease in the merging time window gives a clear improvement in terms of the energy measurement linearity, however a significant saturation effect is retained.

\begin{figure}[]
\centering

\includegraphics[width=0.9\linewidth]{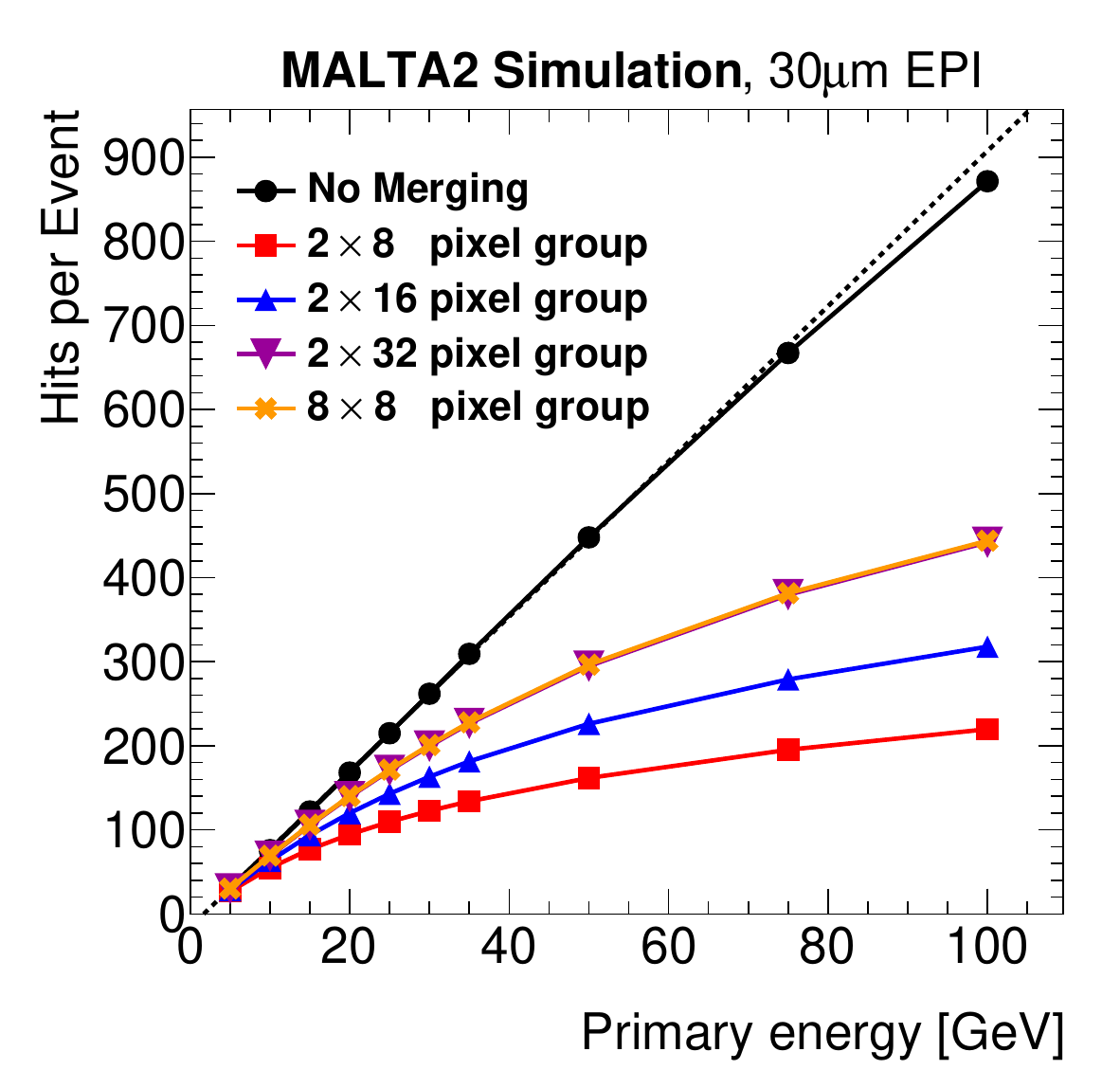}
\caption{Simulation of detected number of hits for an incoming electron of variable energy that interacts upon a \SI{3.2}{cm} tungsten block. 
The different curves are related to various pixel groupings in the digital read-out of the sensor. 
The $2\times8$ group size corresponds to the digital read-out in the real device, while other data sets correspond to possible sensor redesign choices. 
Additionally, the idealized case of no merging is plotted. 
The black dotted line corresponds to a linear fit of the no merging data, performed for electron energies up to \SI{35}{GeV}. 
All simulation points correspond to a \SI{1.6}{ns} merging window.}\label{fig:CalovsGroupSize_UPDATED}

\end{figure}

The effect of the pixel group size on the simulated number of detected hits is investigated in Figure \ref{fig:CalovsGroupSize_UPDATED}. 
A recovery of the saturation effect can be induced with the increase in the pixel group size. 
In comparison to the tracking application investigated in Figure \ref{fig:EffThr_GroupSize_UPDATED}, the orientation of the pixel grouping does not impact the energy reconstruction. 
This is highlighted in the $8\times8$ (orange crosses and interpolated line) pixel group arrangement that perfectly matches the $2\times32$ (purple triangles and interpolated line) arrangement. 
This shows that the merging effect for calorimetry applications is only proportional to the number of pixel group bits. 
On average the increase of the pixel group size has a larger impact on the counting of secondary particles.

\begin{figure}[]
\centering

\includegraphics[width=0.9\linewidth]{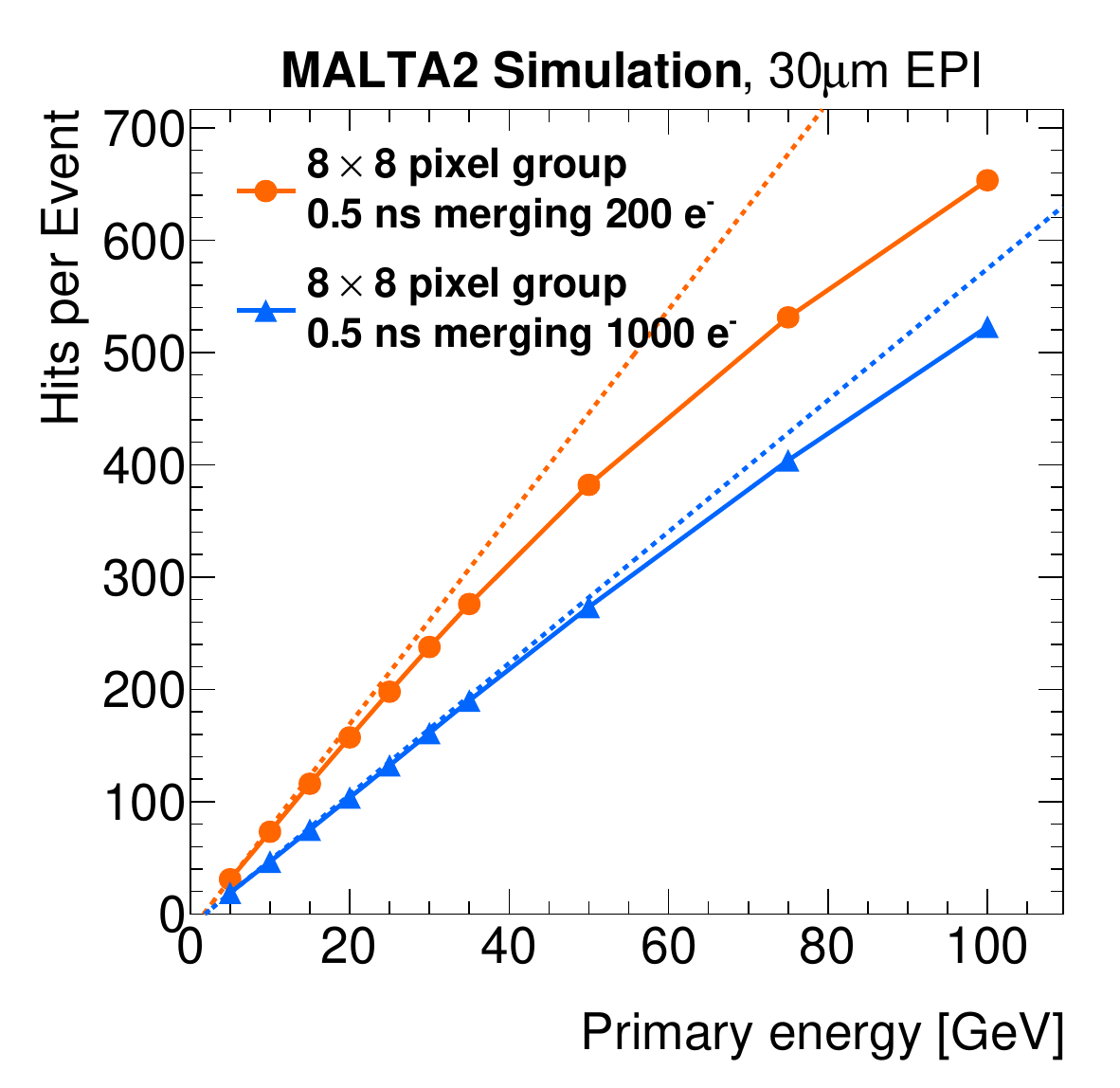}
\caption{Simulation of detected number of hits for an incoming electron of variable energy that interacts with a \SI{3.2}{cm} tungsten block. 
Two curves are plotted corresponding to a $8\times8$ pixel grouping and a merging window of \SI{0.5}{\nano\second}. 
The orange points and interpolated line correspond to a threshold setting of \SI{200}{\mathrm{e^-}}  and the blue triangles and interpolated line to \SI{1000}{\mathrm{e^-}}. 
The dotted lines represent the linear fit of the respective ideal (no merging) data points.}\label{fig:CaloOptimalvsThreshold}

\end{figure}

Combining the lessons learned from the previous investigations, an optimized design can be proposed: an $8\times8$ pixel grouping with a merging window of \SI{0.5}{\nano\second}. 
Figure \ref{fig:CaloOptimalvsThreshold} shows the simulated hit counting linearity for two threshold configurations. 
The nominal threshold (black line and points) configuration (\SI{200}{\mathrm{e}^-}) achieves a good linearity to energies up to \SI{25}{GeV}. 
A further increase in the linearity, up to \SI{50}{GeV} can be achieved by increasing the threshold to \SI{1000}{\mathrm{e}^-}. 
This is due to the smaller number of recorded events, leading to less frequent merging. 
While a better linearity can be achieved at high energies with an increased threshold, a lower energy resolution is expected at lower energies due to the decreased statistics. 
The calorimetry performance can be extended to higher energies by implementing a threshold dependent parametrization. 
A systematic study of both of these effects for a Si-W calorimeter setup with multiple silicon planes is planned to be performed and published in a future paper.

\section{Conclusion}

A data driven, parametric simulation tool for MAPS-based detectors was developed and presented. The parametrization of the charge sharing was deduced with the help of test beam data on the MALTA2 sensor at the SPS beam line. 
The simulation has been validated quantitatively and qualitatively in terms of tracking figures of merit: efficiency and average cluster size. 
An average relative residual between data and simulation of \SI{0.005}{\percent} was measured for the  efficiency and \SI{1.67}{\percent} for the cluster size. 
The hit timing has been validated only qualitatively, showing a reasonably good match between data and simulation. 

The simulation approach allows for a fast and computationally efficient sensor digital design investigation. 
The main source of inefficiencies in the MALTA2 sensor for both tracking and calorimetry applications was found to be the merging circuit. 
Improvements for both applications were investigated in terms of silicon sensor redesigns focused on the digital domain.

The leading order improvements on the efficiency have been found to be a decreased merging time window. 
A secondary order impact was found to be the increase in the pixel group, with a preference for a grouping that increases the area to perimeter ratio. 
A mix of such sensor redesigns is expected to increase the efficiency, above a value of \SI{99.8}{\percent}.

In terms of digital calorimetry, it has been shown that following minor digital redesigns, the MALTA sensor retains a linear hit counting for electromagnetic showers produced by electrons with energies up to \SI{50}{GeV}. 
The electromagnetic showers correspond to the shower maximum, giving a worst case calorimetry estimation. 
A multi-plane calorimeter expects a better linearity up to higher energies. 
A systematic study of a multi-layer, fully digital calorimeter for high energy physics is planned in a future work. 

The largest improvements in energy reconstruction linearity was found to come from the size of the pixel group. 
No preference in terms of the pixel group arrangement was found. 
Additional improvements on the counting linearity were found to be a smaller merging window and larger thresholds. 
An optimized design for future calorimetry applications would include a mix  or all of these improvements, depending on the physics case. 

The successful simulation of MALTA2 sensors enables future large scale detector simulation. 
A complementary TCAD simulation of MALTA2 sensors is also pursued within the collaboration.
Additionally, the implementation of custom read-out circuitry in the simulation is expected to guide future sensor design for tracking and digital calorimetry applications.

\bmhead{Acknowledgements}
This project has received funding from the European Union’s Horizon 2020 Research and Innovation programme under Grant Agreement number 101004761 (AIDAinnova), and number 654168 (IJS, Ljubljana, Slovenia). 
Furthermore it has been supported by the Marie Sklodowska-Curie Innovative Training Network of the European Commission Horizon 2020 Programme under contract number 675587 (STREAM).

\bibliography{epj}

\backmatter

\end{document}